\def\simlt{\mathrel{\rlap{\lower 3pt\hbox{$\sim$}}\raise
2.0pt\hbox{$<$}}}
\def\simgt{\mathrel{\rlap{\lower 3pt\hbox{$\sim$}} \raise
2.0pt\hbox{$>$}}}

\newcommand{\q}{\begin{equation}}
\newcommand{\qa}{\begin{eqnarray}}
\newcommand{\qs}{\begin{eqnarray*}}
\newcommand{\nq}{\end{equation}}
\newcommand{\nqa}{\end{eqnarray}}
\newcommand{\nqs}{\end{eqnarray*}}
\def\be{\begin{equation}}
\def\ee{\end{equation}}

\def\etal{{\it et al.~}}

\documentclass[useAMS,usenatbib,usegraphicx]{mn2e}

\def\aap{A\&A }

\def\apj{ApJ }

\def\mnras{MNRAS }
\newcommand{\gsim}{\raisebox{-3.8pt}{$\;\stackrel{\textstyle >}{\sim}\;$}}
\newcommand{\lsim}{\raisebox{-3.8pt}{$\;\stackrel{\textstyle <}{\sim}\;$}}

\title[Simulation of clusters--I]{Simulating galaxy clusters -- I. Thermal and chemical properties of
the intra-cluster medium}
\author[Romeo et~al.]{
A. D. Romeo$^{1,2,3}$
\thanks{E-mail: aro@na.astro.it}, J. Sommer-Larsen$^{1,4}$\thanks{E-mail: jslarsen@tac.dk}, 
L. Portinari$^{1,5}$\thanks{E-mail: lporti@utu.fi}, V. Antonuccio-Delogu$^{6}$\thanks{E-mail:
van@ct.astro.it}\\
$^{1}$ Teoretisk Astrofysik Center, Juliane Maries Vej 30, DK-2100 Copenhagen \O, DENMARK\\
$^{2}$ Dipartimento di Fisica e Astronomia, Universit\`{a} di Catania, via S.Sofia 64, I-95123 Catania, ITALY\\
$^{3}$ INAF--Osservatorio Astronomico di Capodimonte Napoli, salita Moiariello 16, I-80131 Naples, ITALY\\
$^{4}$ Dark Cosmology Centre, Niels Bohr Institute, Juliane Maries Vej 30, DK-2100 Copenhagen \O, DENMARK\\
$^{5}$ Tuorla Observatory, V\"ais\"al\"antie 20, FIN-21500 Piikki\"o, FINLAND\\
$^{6}$ INAF--Osservatorio Astrofisico di Catania, via S.Sofia 78, I-95123 Catania, ITALY\\}

\begin{document}

\date{Accepted ?. Received ?; in original form ?}

\pagerange{\pageref{firstpage}--\pageref{lastpage}} \pubyear{2006}

\maketitle

\label{firstpage}
\begin{abstract}
We have performed a series of N-body/hydrodynamical (TreeSPH) simulations of 
clusters and groups of galaxies, selected from a cosmological 
volume within a $\Lambda$CDM framework: these objects have been
re-simulated at higher resolution to $z$=0, in order to follow also the 
dynamical, thermal and chemical input on to the ICM from stellar populations 
within galaxies.
The simulations include 
metallicity dependent radiative cooling,
star formation according to different IMFs, 
energy feedback as strong starburst-driven galactic super-winds,
chemical evolution with non-instantaneous recycling of gas and heavy elements, 
effects of a meta-galactic UV field and thermal conduction in the ICM.

In this Paper I of a series of three, 
we derive results, mainly at $z=0$, on the temperature and entropy 
profiles of the ICM, its X-ray luminosity, the cluster cold components (cold fraction 
as well as mass--to--light ratio) and the metal distribution between ICM and stars.

In general, models with efficient super-winds (produced by the action 
of supernov\ae\ and, in some simulations, of AGNs), along with a top-heavy stellar IMF, are
able to reproduce fairly well the observed $L_X-T$ relation, the entropy profiles
and the cold fraction: both features are found to be needed in order to remove
high-density and low-entropy cold gas at core scales, although additional alternative
feedback mechanisms would still be required to prevent late time central
cooling flows, and subsequent overproduction of stars and
heavy elements at the centre. 
Observed radial ICM temperature profiles can be matched, except for the
gradual decline in temperature inside of $r\sim$~0.1$R_{\rm{vir}}$.
Metal enrichment of the
ICM gives rise to somewhat steep inner iron gradients; 
yet, the global level of enrichment compares well to observational estimates
when a top--heavy IMF is adopted, and after correcting for
the stars formed at late times at the base of the cooling flows, the metal
partition between stars and ICM gets into good agreement with 
observations. 
The overall abundance and profile of iron in the ICM is found
essentially unchanged from $z=1$ to present time. 
Finally, the $\frac{\alpha}{Fe}$ of the gas is found to increase steadily 
with radius, decreasing over time.
\end{abstract}

\section{Introduction}
Clusters of galaxies play a major role in most of the attempts at
reconstructing the cosmological history of the Universe. Numerical
simulations (since Evrard 1990, and then Thomas \& Couchman 1992, Katz \& 
White 1993, Cen \& Ostriker 1994, Navarro \etal 1995, Evrard \etal 1996,
Eke \etal 1998, Thomas \etal 2002, Tornatore \etal 2003, Borgani \etal 2004) are 
a powerful tool for
making predictions on their formation and evolution, provided that the
underlying physical mechanisms are fully taken into 
account. 

In particular cosmological simulations have helped to reproduce
many of the X-ray properties of the intra--cluster medium (ICM), which
is a hot (1-10 keV) tenuous ($10^{-4} - 10^{-2}$ cm$^{-3}$), optically thin
gas containing
completely ionized H and He and partially ionized heavier elements ( ``metals"). 
Metal lines revealed to populate spectra of both groups and clusters, corresponding 
to metallicities of about 0.2 to 0.4 in solar units, which indicates
a non purely primordial origin of the ICM, a fraction of which must have previously 
been processed in galaxies as inter--stellar medium (ISM) and then
transported whence into the space between galaxies. 

The ICM is emitting radiation at X-ray wavelenghts by losing energy 
according to a cooling timescale $t_{cool}\propto \sqrt{T}/n$. 
In the very central region of many clusters
the gas density becomes high enough to make the cooling time shorter
than a Hubble time, forming thus a cooler core (typically 
$R_c \sim$100 kpc or more) 
which affects the gas central density, temperature
and entropy distributions (see Sections 3, 4, 5).
The ICM has likely been heated up to X-ray emitting temperatures mainly 
by means of gravitational 
energy converted to thermal when falling onto the cluster's potential well. 
Indeed the basic scenario describing the thermal behaviour of the ICM calls
for a gravity-only driven ongoing heating of the gas by adiabatic compression,
followed by hydrodynamical shocks arisen during the settling of infalling 
material. The hot gas will emit mainly by thermal (free-free)
Bremsstrahlung, resulting in 
the simple, theoretical clusters 
X-ray scaling relations: luminosity--temperature $L_X\propto T^2(1+z)^{3/2}$,
entropy--temperature $S\propto T(1+z)^{-2}$, virial mass--temperature 
$M\propto T^{3/2}$.

It is by now well known how this simple model basically fails in reproducing
the actual scaling properties observed: in fact, the $L_X-T$ relation at $z=0$
is steeper than predicted, breaking possibly into two different slopes (see Section 6).
As for the S-T relation, an entropy excess with respect to the self-similar 
expectation has been reported in the inner region of
poorer and cooler systems, reaching a constant value (``entropy floor"), 
likely resulting from softer gas density profiles in the core (Section 5). 
Finally the M-T relation 
has been found to have a lower normalization than predicted by ``adiabatic" models 
and simulations of up to $\sim 40\%$.

Physical mechanisms able to explain the break of self--similarity in the
properties of the ICM have been extensively proposed during last ten years, and
mainly fall in two broad classes: a) non--gravitational heating and
b) radiative cooling. Introducing gas cooling and star formation induces
gas condensation at core scales, leading to star concentration 
associated with
the central galaxy (cD) which will be dominating the baryonic cluster mass in the 
core at $z=0$. This affects also the dark-matter (DM) density profile in the
core, which steepens from $r^{-1.5}$ to $r^{-2}$. Moreover, inflows of high--entropy 
gas from extra--core regions are stimulated by virtue of baryon
cooling, while lower-entropy, shorter-cooling time gas gets selectively
removed from the centre, thus giving rise to an increased level of entropy 
inside the core itself (Voit \& Bryan 2001). On the other hand X-ray luminosity 
is suppressed
since the amount of gas populating the hot diffuse phase gets reduced (Pearce 
\etal, 2000). However all cluster simulations with cooling and star formation
reveal a significant drawback: cooling alone makes a large fraction (30 up
to 55\%) of gas get converted into a cold phase (either cold gas or stars)
within the virial radius at $z=0$ (Suginohara \& Ostriker 1998, Balogh \etal 2001, 
Dav\`e \etal 2001), 
in contradiction with a value of 10-20\% of the baryon fraction 
locked into stars indicated by observations. 
 
Thus, additional sources
of non-gravitational heating are required to counterbalance over--cooling 
in the cluster's core,
to enhance the entropy of the gas and to suppress its X-ray emissivity 
(Evrard \& Henry 1991, Bower 1997, 
Lewis \etal 2000, Bialek \etal 2001). 
One solution would be to appeal to pre--heating, namely
episodic mechanisms such as an impulsive injection of energy to all gas particles 
at redshift high enough to occur either before (external pre--heating:
requires less energy to reach a given entropic level because occurring in a
less dense pristine environment: e.g. Kaiser 1991, Cavaliere \etal 
1997, Balogh \etal 1999, Mo \& Mao 2002), or during the gravitational collapse
onto the cluster's potential, or shock heating during late
accretion within DM haloes (internal pre--heating: e.g. Tozzi \& Norman 2001, 
Babul \etal 2002, Voit \etal 2003): its effect would be 
clearly stronger for colder systems (groups and small clusters), where both
excess of inner entropy and a steeper L-T relation can be achieved 
(Cavaliere \etal 1998, Brighenti \& Mathews 2001, Borgani \etal 2001). 

Feedback processes subsequent to star formation, namely supernov\ae~induced 
galactic winds, are natural mechanisms working at both pre--heating
and especially metal enriching the ICM (Loewenstein \& Mushotzky 1996, Renzini 1997, 
Finoguenov \& Ponman 1999, Wu \etal 2000, Lloyd-Davies \etal 2000, Kravtsov \& Yepes 
2000, Menci \& Cavaliere 2000, Bower \etal 2001, Pipino \etal 2002, Kapferer
\etal 2006);
yet doubts have been raised as to
the actual efficiency in thermalizing the amount of additional supernova energy,
which would be maximized e.g. in starbursts. 
Domainko\ etal (2004) took into account the effect of heating and metal enrichment 
from intra-cluster SNe, the energy from which may inhibit
or delay the formation of cooling flows, provided that the inter-galactic star fraction
is as considerably high as shown in other simulations ($\gsim$20-25\%, see Sommer-Larsen, Romeo
and Portinari, 2005, hereinafter Paper III).
Up to now the most promising candidates to supply energy for heating are AGNs
associated with quasars, which can have at their disposal huge energetic reservoirs 
to be converted into gas thermal energy (Bower 1997, Valageas \& Silk 1999ab, 
Wu \etal 2000, McNamara \etal 2000, Yamada \& Fujita 2001; Cavaliere, Lapi \&
Menci 2002, Croton \etal 2006).
In conclusion, stellar feedback and/or additional pre--heating must play
together with gas cooling to regulate the latter such as to suppress
over--cooling (see Pearce \etal 2000, Kay \etal 2003): this will set up the fraction 
of baryons cooled out (``cold fraction") to form stars and cold clouds, 
which in turn drives
the distribution of gas density, temperature and entropy.

So far few hydrodynamical simulations have been performed which were able
to pursue at the same time and in a self--consistent way the target of
realistically modelling the ICM dynamics by including both non-gravitational
heating/cooling (complete with feedback/star formation) {\it and} the metal
enrichment associated with star formation and stellar evolution:
Borgani \etal (2001 and 2002), Tornatore \etal (2003), McCarthy \etal (2004)
(all though without
following metal production from star formation), and especially Valdarnini
(2003), Tornatore \etal (2004) and, very  recently, Scannapieco \etal (2005)
and Schindler \etal (2005).

The physical processes implemented in our simulations and the numerical details are 
described in Section 2. In Section 3 to 6 we discuss distribution and scaling of 
density, temperature, entropy and luminosity of the X--ray emitting gas. 
In Section 7 and 8 we discuss the relation of the hot gas with the cold stellar 
component and the consequent chemical enrichment. Finally, in Section 9 we draw 
some summary and conclusions.

\section{The Code and Simulations}
The cosmological N-body simulation has been performed using the FLY code 
(Antonuccio \etal, 2003) for a
$\Lambda$CDM model ($\Omega_0$=0.3, $\Omega_{\Lambda}$=0.7, $\Omega_b$=0.04, 
$h=0.7$, $\sigma_8$=0.9) with $128^3$ collisionless DM-only particles filling 
an initial box of 150$h^{-1}$ Mpc at an initial redshift $z_i=39$. 
The gravitationally
bound DM haloes at $z$=0 were identified by a FOF group-finder, with a fixed 
linking length (at $z$=0) equal to 0.2 times the average
initial interparticle distance (see e.g. Eisenstein \& Hut 1998).

Two small groups, two large groups and two clusters were sorted out from the cosmological simulation, 
resampled to higher resolution and resimulated by means of an improved version of the Tree-SPH 
code described in Sommer-Larsen, G\"otz \& Portinari (2003; hereinafter
SLGP). We named the two clusters after ``Virgo'' and ``Coma'', because of their virial size:
$\sim 3$ and $\sim 6$ keV in terms of X-ray temperature (see table 1), respectively
(actually ``Coma'' rather resembles a ``mini-Coma''-like system, being the real Coma at about 8 keV).
For each object, all DM particles within
the virial radius at $z=0$ have been traced back to the initial condition at $z_{i}$,
where the resolution in this ``virial volume'' has been increased (see below)
and one SPH particle has been added per each DM one, with a mass of $m_{SPH}=
f_b\cdot m_{DM}^0$, resulting in a reduced dark-matter mass $m_{DM}=(1-f_b)\cdot
m_{DM}^0$: here and throughout this series of papers we have used a baryon
mass fraction $f_b=\frac{m_{SPH}}{m_{DM}+m_{SPH}}=0.12$, generally consistent 
with $\Omega_b h^2$ from nucleosynthesis and cluster
observational constraints. The mass resolution reached inside the resampled
lagrangian subvolumes from the cosmological set was such that 
$m_{DM}=1.8\cdot 10^9 h^{-1} M_{\odot}$
and $m_{g}=m_*=2.5\cdot 10^8 h^{-1} M_{\odot}$ (the mass of star
particles equals that of SPH gas particles throughout the simulation, see
Section~2.3). In one high-resolution run of the ``Virgo'' cluster, it was
$m_{DM}=2.3\cdot 10^8 h^{-1} M_{\odot}$ and
$m_{g}=m_*=3.1\cdot 10^7 h^{-1} M_{\odot}$. In the surrounding region of these
subvolumes the DM particles were increasingly resampled at coarser resolution with 
increasing radial distance (see Gelato \& Sommer-Larsen, 1999), leaving a buffer 
layer in between at intermediate resolution. 
As a final configuration, we have
therefore an inner region corresponding to the initial virial volume and
containing the highest-resolution (64 times the basic value; 512 times
for the high resolution run) DM and SPH particles, 
a buffer region of DM only particles at 8 times resolution, and
finally an outer envelope of DM only particles at base resolution.

Softening lenghts controlling gravitational interactions among
particles are $\epsilon_{DM}=5.4 h^{-1}$kpc and $\epsilon_{g}= 
\epsilon_{*}=2.8 h^{-1}$
kpc (for the high-resolution run: 2.7 and 1.4 $h^{-1}$kpc, respectively); 
they were kept constant in physical
space (meaning increasing with z in comoving coordinates) since $z=6$
and conversely fixed in comoving coordinates at earlier times.
These values of the softening lengths for the DM particles correspond to a
virial mass of $3.66 \cdot 10^7 h^2 M_{\odot}$, about two orders of magnitude 
less than
the minimum mass of our particles, which ensures that the gravitational
potential is well reproduced even within gravitationally bound groups.


The selected objects' characteristics are listed in Tables 1 and 2;
they have all been chosen such as not to be undergoing any major merger events
since $z\sim 1$.\\
The adopted TreeSPH code schematically includes the following features (see Table 3):

\begin{itemize}
\item[$\bullet$] solution of entropy equation (rather than of thermal energy)
\item[$\bullet$] metal-dependent atomic radiative cooling
\item[$\bullet$] star formation
\item[$\bullet$] feedback by starburst (and optionally AGN) driven galactic winds
\item[$\bullet$] chemical evolution with non-instantaneous recycling of gas and heavy
elements (H, He, C, N, O, Mg, Si, S, Ca, Fe)
\item[$\bullet$] thermal conduction
\item[$\bullet$] meta-galactic, redshift-dependent UV field. 
\end{itemize}

In the following subsections we describe in
more details how star formation and feedback mechanisms work, and how we
implement chemical evolution with cooling.

\subsection{Star Formation and Feedback as Super-Winds}

Although only a minor fraction of the baryons in groups and clusters is in
the form of stars and cold gas, the ICM is significantly enriched in heavy 
elements, with an iron abundance about 1/3 solar in iron (e.g.\ Arnaud 
\etal 2001; De Grandi \& Molendi 2001; De Grandi \etal 2004; 
Tamura \etal 2004, to quote a few recent {\it BeppoSAX} and {\it XMM} 
studies).
In fact one can show from quite general arguments that at least half and
probably more 
of the iron in galaxy clusters resides in the hot ICM with most of the
remaining iron locked up in stars (Renzini \etal 1993;
Renzini 1997, 2004; Portinari \etal 2004). Moreover, 
observations of galaxy clusters up to $z$$\sim$1 indicate that the ICM 
metals were already widely in place at this redshift (Tozzi \etal 2003).

Since the metals must have mainly or exclusively been manufactered in 
galaxies (whether still existing or by now disrupted into an intra--cluster 
stellar population), efficient gas and metals outflows from these 
must have been substantially at work in the past ($z$$\ga$1): 
we shall in the following denote these ``galactic super-winds", and
SW simulations those run with such strong wind feedback.
The galactic super-winds may be driven by supernov\ae\ or hypernovae (e.g. Larson 
1974, Dekel \& Silk 1986, Mori \etal 1997, Mac Low \& Ferrara 1999, SLGP), AGNs 
(e.g., Silk \& Rees 1998; Ciotti \& Ostriker 1997, 2001; Romano 
{\it et al.} 2002; Springel, Di Matteo \& Hernquist 2005; 
Zanni {\it et al.}\ 2005), or a combination of the two.

Recent work on galaxy formation indicates that in order to get a significant
population of disc galaxies with realistic properties in addition to
steady, quasi self-regulated star formation, one has to invoke 
early non-equilibrium outflow processes (e.g., Abadi \etal 2003, SLGP, Governato 
\etal 2004, Robertson \etal 2004). 
Such early star-burst driven outflows were incorporated in the simulations
of SLGP, and in the present simulations we build on
this approach, except that in general we allow such outflows to occur at
all times (all ``SW" runs: see below).

Specifically our approach is the following: cold gas particles 
($T$$<$$2\cdot 10^4$K) are triggered for star formation on a timescale
$t_*=t_{dyn}/\epsilon_{SF}$ (where the star formation efficiency 
$\epsilon_{SF}=0.02$) if the gas density 
exceeds a certain critical value, chosen to be $n_{\rm{H}}^{up}=1.0$ cm$^{-3}$;
we have experimented with other values and found that the outcome of the
simulations is very robust with respect to changes in this parameter. To model the 
sub-resolution physics of star-bursts and self propagating star formation in a
simplistic way we assume that, if a SPH particle satisfies the above criterion
and has been triggered for star formation, then with a probability 
$p_{\rm{burst}}$ its neighbouring
cold and dense SPH particles with densities above $n_{\rm{H}}^{low}
(< n_{\rm{H}}^{up})$ are also triggered for star formation on their respective
dynamical time scales. The fraction $f_{\rm{wind}}$ of all stars 
formed which partake in bursts is controlled through the parameters 
$n_{\rm{H}}^{low}$ and $p_{\rm{burst}}$. Simulations with combinations of
the latter two parameters yielding similar values of $f_{\rm{wind}}$ give
similar results, so effectively the wind ``strength" is controlled by one
parameter. For most simulations presented in this paper we have used 
$n_{\rm{H}}^{low}=0.8$ cm$^{-3}$ and $p_{\rm{burst}}$=0.1, resulting in
$f_{\rm{wind}}$$\simeq$0.8.

When a star particle is born, it is assumed to represent a population of 
stars formed at 
the same time in accordance with a Salpeter or an Arimoto-Yoshii (Arimoto \& Yoshii, 1987) IMF. 
It feeds energy back to the local ISM by Type II supernov\ae\ (SNII) explosions 
during the period 3-34 Myr after formation: in fact, the lightest stars which 
contribute feedback from SNII explosion have a mass of about 9 $M_{\odot}$ 
and lifetimes of about 34 Myr; 
and the upper limit of the IMF is assumed to be 100 $M_{\odot}$ corresponding to
a lifetime of about 3 Myr (see Lia, Portinari \& Carraro 2002). 
Stars with masses greater than 9 $M_{\odot}$ are assumed to deposit $\sim$
10$^{51}$ ergs per star to the ISM as they explode as SNII.
Throughout, the energy output from stellar winds is
neglected, because even for metal-rich stars it is an order of magnitude
less than the output from SNII.\footnote{Stars of $6-9~M_{\odot}$ also produce SN~II, 
mostly via the electron--capture mechanism (Portinari, Chiosi \& Bressan
1998); however, due to the longer lifetime of the progenitors (34--75 Myr), 
these explosions are much more diluted in time and less correlated in space, 
giving no burst and super--shell effects. We thus neglect energy
feedback from this mass range, also because the endpoint of the evolution
is still debated: Super--Asymptotic Giant Branch evolution might develop 
in the non--violent formation of a massive white dwarf, or
in a weak explosion at most (Ritossa, Garcia-Berro \& Iben 1996, 1999;
Eldridge \& Tout 2004).}

SNII events are associated with multiple explosions of coeval stars in
young stellar (super-)clusters,
hence resulting in very efficient contribution to energy feedback.
In the simulations, the energy from SN~II is deposited into the ISM as thermal energy 
at a constant rate during
the aforementioned period (the energy is fed back to the, at any time, 50 nearest 
SPH particles using the smoothing kernel of Monaghan \& Lattanzio 1985). Part 
of this thermal energy is subsequently converted by the code into kinetic energy 
as the resulting shock-front expands. While a star
particle is feeding energy back to its neighbouring SPH particles, radiative 
cooling of these is switched off: this is an 
effective way of modelling with SPH a two-phase 
ISM consisting of a hot component ($T \sim 10^6-10^7$ K) and a much cooler
component ($T \sim 10^4$ K) --- see Mori \etal \cite{M.97}, Gerritsen 
\cite{G97} and Thacker \& Couchman (2000 and 2001). At the same time 
interrupting cooling in the surrounding region of star particles undergoing SNII
explosion, helps mimicking the adiabatic phase of expansion, thus creating 
super-shells. Such an adiabatic phase of supernov\ae\ explosion 
would be missed in the simulations if the whole thermal energy released were to get 
dissipated by efficient cooling mechanisms.
The feedback related to the (non star-burst) conversion of individual, 
isolated SPH particles to star particles typically results in 
``fountain-like'' features, and does not lead to complete ``blow-away''.

In fact, a typical star-burst event as described above gives origin to 40-50 
star particles which 
are at formation localized in space and time. The energy feedback from these star  
particles drives strong shocks/``super-shells'' outwards through the 
surrounding gas, sweeping up gas and metals and transporting these to (and
mixing into) the 
ICM --- such outflows being the ``galactic super-winds'' we introduced. These 
super-winds are quite well resolved numerically --- a further improvement in 
the present work has been the adoption of the ``conservative'' 
entropy equation solving scheme of 
Springel \& Hernquist (2002), which improves shock resolution
over classic SPH codes.

The amount of energy injected into the Inter Stellar Medium (ISM) by Type~II 
supernovae per unit mass of stars initially formed can be expressed as   
$\propto f_{\rm{AGN}}~\beta~\nu_{SN}~E_{SN}$,
where $f_{\rm{AGN}}$ normally is unity (but see below), $\beta \le 1$ accounts
for possible radiative losses, $\nu_{SN}$ is
the number of Type~II supernovae produced per unit mass of stars formed and
$E_{SN}$ is the amount of energy released per supernova. For this work we
assume $\beta E_{SN}$=10$^{51}$~ergs. The parameter $\nu_{SN}$ depends on
the IMF: for the Salpeter IMF $\nu_{SN}=6 \times 10^{-3}$, for the 
Arimoto-Yoshii IMF $\nu_{SN}=1.5 \times 10^{-2}$, corresponding to 6 and
15 SNII respectively (from progenitors with $M \geq 9~M_{\odot}$), per 
1000~M$_{\odot}$ of stars formed.
To mimic the effect of possible additional feedback from central,
super-massive black holes in a very crude way we set in some simulations
$f_{\rm{AGN}}$=2 or 4 (SWx2 and SWx4 runs respectively), on the basis 
of the hypothesis
that AGN activity mirror star formation (Boyle \& Terlevich 1998). 

As a test of reference, one set of simulations (``WFB") were run with
early super-winds only, as SLGP assumed for modelling individual 
galaxies. This resulted
in a very small (average) value of $f_{\rm{wind}}$=0.07, and a consequent
very low level of enrichment of the ICM. As a result, this WFB set has been run 
using an approximately primordial cooling function in the ICM, 
and hence is similar to most of the previously attempted simulations 
performed without metal production from star formation and subsequent chemical 
evolution of the ICM ({\it e. g.} Tornatore \etal, 2003).

The above ``super-wind'' scheme is not resolution independent by construction,
yet simulations at 8 times higher mass and twice better force resolution
give similar results (see {\it e.g.} Section 6 and also Paper II), 
indicating that the scheme is effectively 
resolution independent. 
Although the above scheme 
is just a simplistic way of modeling galactic outflows, it has the advantage
that such outflows are numerically resolved, and that the gas dynamics of 
the outflow process is self-consistently handled by the code.

Finally, Type Ia supernov\ae\ (SNIa) are also implemented in the simulations as 
sources
of chemical enrichment. However,since SNIa explosions are considerably less
numerous than SNII (per unit mass of stars initially formed), and also  
are typically uncorrelated in space and time, we neglect their energy
feedback in the current simulations. As a matter of fact, SNIa explosions,
occurring in low-density environments, may be in principle more efficient in
contributing to feedback (Recchi, Matteucci \& D'Ercole 2001; Pipino et~al.\ 2002); 
nonetheless they represent isolated events not
giving rise to cumulative, large scale effects such as the supershells 
resulting from correlated SNII explosions.


\begin{table}
\caption{Structural properties of selected clusters and groups at $z$=0
\label{t:1}}
\small
\begin{tabular}{lccc}
\hline\hline
Cluster \#  & $M_{vir}$ & $R_{vir}$ & $<kT_{ew}>$ $^{\rm a}$ \\
  &  [10$^{14}$M$_{\odot}$] & [Mpc] &[keV]~\\
\hline
``Coma" & 12.38 &2.90 & 6.00\\
``Virgo" &  2.77 & 1.77 & 3.06\\
202 &  1.25 & 1.29 & 2.24\\
215 &  1.03 & 1.18 & 1.48\\
550 &  0.48 & 0.96 & 1.00\\
563 &  0.49 & 0.94 & 1.10\\
\hline \hline
\multicolumn{4}{l}{$^{\rm a}$ Temperatures referred to the "standard" run AY-SW}\\ 
\end{tabular}
\end{table}

\begin{table}
\caption{Numerical properties of the simulations. All the runs were started 
from initial redshift $z_i$=19, except ``Virgo'' high-resolution, for which
$z_i$=39; throughout the paper, all results from the latter refer to $z=0.07$,
up to which the simulation has actually run as of submission date.
Last two columns refer to SW runs.
\label{t:1}}
\small
\begin{tabular}{lrrrr}
\hline\hline
Cluster \#  & $N_{DM}$ & $N_{SPH}$ & $N_{DM}^{z=0}$ & $N_{SPH+*}^{z=0}$ \\
	& & & $(r<R_{vir})$ & $(r<R_{vir})$ \\
\hline
``Coma"  & 517899 & 429885 & 356300 & 330850\\
``Virgo" & 148634 & 109974 & 82461 & 75918 \\
``Virgo''x8 & 1140731 & 1093992 & 707500 & 646547\\
202 &  69330 & 52192 & 40900 & 36230\\
215 &  62265 & 46294 & 30950 & 27490\\
550 &  51392 & 33734 & 16875 & 13667\\
563 &  44872 & 28056 & 15800 & 13760\\
\hline \hline
\end{tabular}
\end{table}

\begin{table}
\caption{Characteristics of the runs: all implement the SW scheme 
($\epsilon_{SF}=0.02$, $f_{wind}=0.8$), except Sal-WFB and ADIAB
\label{t:2}}
\small
\begin{tabular}{lcccc}
\hline\hline
\# & $IMF$ & $f_{AGN}$ & $f_{wind}$ & preh.@ z=3 \\
 & & & & [keV/part.]\\
\hline
AY-SW & AY & 1 & 0.8 & 0 \\
AY-SWx2 & AY & 2 & 0.8 & 0 \\
AY-SWx4 & AY & 4 & 0.8 & 0 \\
PH0.75 & AY & 1 & 0.8 & 0.75 \\
PH1.50 & AY & 1 & 0.8 & 1.50 \\
PH50 & AY & 1 & 0.8 & 50 $\cdot$ cm$^2$ \\
Sal-SW & Sal & 1 & 0.8 & 0 \\
Sal-WFB & Sal & 1 & 0.07 & 0 \\
COND & AY & 1 & 0.8 & 0 \\
ADIAB & -- & -- & -- & 0 \\
\hline \hline
\end{tabular}
\end{table}

\subsection{Cooling flow correction}

The simulations show active star formation, induced by strong 
cooling flows, at the very centre of the cD galaxies 
down to $z=0$, which has no counterpart in observed clusters; 
such spurious late activity is responsible for
a non negligible fraction of the cluster stellar mass and 
these young stellar populations would significantly 
contribute to the total luminosity. Therefore a correction 
must be applied, if aiming at reproducing 
observed quantities such as cold fraction, star formation rate  
as well as the metal production and distribution in the ICM 
(see Sections 7 and 8; also Paper III and D'Onghia {\it et al.} 2005). 

Specifically, our correction consists
in selecting and subtracting out the star particles formed within
the innermost 10~kpc since some $z_{corr}$:
as the main epoch of general galaxy formation in the clusters
is over by $z \sim 2$ (see Paper II), we assume $z_{corr}$=2
as the redshift below which spurious central star formation induced
by cooling flows can be neglected;
we also verified that the correction is quite stable for $z_{corr}$
up to 2.5.

\subsection{Chemical evolution}

There is a deep interconnection between chemical and hydrodynamical
evolution when addressing metal production and mixing into the ICM.
Stars return to the surrounding ISM part of their mass, including chemically
processed gas, and of their energy (stellar feedback):
the whole star formation process acts onto the cluster gas evolution
both through energy feedback from supernov\ae\ (see previous section), 
counteracting the cool-out of the hot gas,
and through metal enrichment of the ICM, which instead boosts its cooling 
rate. 
The rates of release of both energy (SN rate) and metals are set by
the IMF, which also regulates the returned fraction from the stellar to the
gaseous phase: in a Salpeter distribution, 30\% of the stellar mass formed
is returned to the gaseous phase within a Hubble time, whilst in an 
Arimoto-Yoshii (AY) distribution the returned fraction is about 50\%.

In our simulations, star formation and chemical evolution are modelled
following the ``stochastic'' algorithm of Lia {\it et~al.}\
(2002; hereinafter LPC). When a
cold gas particle is selected for star formation, 
it is entirely transformed to a star
particle with a probability given by the star formation efficiency.
The new-born collisionless star particle hosts a Single Stellar Population
(SSP), namely a distribution of stellar masses (according to the chosen IMF)
all coeval and with homogeneous chemical composition, which is the composition
of the gas particle that switched to a star particle. During its lifetime,
such a SSP will release gas and metals (both newly synthesized and original
ones). The return of gas and metals to the interstellar medium is also
modelled stochastically, and a star particle will entirely ``decay'', 
or ``return'' back
to be a SPH collisional particle, with a probability given by the returned
fraction of the corresponding IMF. Which is to say, in simulations with the
Salpeter IMF 30\% of the star particles will return to be gas in a Hubble
time, while with the AY IMF 50\% of the star particles decay back to gas 
particles, overall. Notice that the decay is not instantaneous: only 10\% (for
Salpeter, 30\% for AY) of the SSP mass is returned to the ISM in the initial
burst phase (the initial 34 Myr, Section 2.1), the rest of the decay of the
star particles occurs spread over a Hubble time. According to its age, 
a decaying
star particle carries along to the gaseous phase the metals produced by the
SSP in that time window (namely, the products of the dying stars and exploding
supernov\ae\ with the corresponding lifetime). Thus the algorithm follows
non--instantanoues gas recycling, as well as the corresponding delayed metal
production. In particular, mass loss and metal production from intermediate
and low mass stars, and the delayed chemical enrichment from SNIa, 
are accounted for.
The algorithm follows the evolution of H, He, C, N, O, Mg, Si, S, Ca and Fe.
Full details of the algorithm can be found in LPC.
Among the advantages of this algorithm, the number of baryonic particles and
their mass is conserved throughout the simulation: since they switch from gas 
to stars, and back from stars to gas, in their entirety with no splitting,
$m_*=m_{SPH}$ always.

Some modifications have been applied to the original LPC
algorithm, to adapt it to our ``starburst and supershell'' feedback scheme
described in the previous section.
In the first 34~Myrs, i.e.\ during the starburst phase, the energy and metals
released from massive stars are not implemented statistically; rather, with the
rates predicted by the corresponding IMF, there is a steady release of
energy and metals over the neighbouring gas particles, hence the expanding 
super-shell is
also enriched in metals. When reaching 34~Myr, a fraction of 
the star particles responsible
for the burst are turned to SPH particles, with the suitable 
probability
to describe the actual returned gas fraction at that age (about 10\% or 30\%
for the Salpeter or AY IMF, see above). Only from 34 Myrs onwards, metal production
is treated with the stochastic approach of LPC. Being limited to the very
early, short stages of the SSP lifetime, this modification does not hamper 
the correct
modelling of delayed gas recycling and metal production achieved with the LPC
algorithm; but it allows a better connection between energy and metal feedback
from SNII in our starburst models, and a more efficient transport of the SNII
metals outward by the expansion of the corresponding supershell.

Further mixing of the metals, both the SNII metals and those released by later 
decay of star particles, could be due to metal diffusion. In LPC, metal
diffusion was taken into account with a diffusion coefficient $\kappa$ based
on the expansion of individual supernova remnants; this is relevant for
modelling galactic scales, but can be neglected on cluster
scales. Besides, in our simulations the diffusion related to the expansion of 
(multiple) SN shells is implicit in the treatment of the winds. Hence in our 
present simulations we neglect (additional) metal diffusion 
($\kappa$=0).

%
\subsection{Background UV field and metal-dependent cooling}

A source of radiative heating is provided by a homogeneous and isotropic UV
background radiation field switched on at $z=6$ and modelled after Haardt \&
Madau (1996), to integrate the effect of sources such as 
AGNs and/or young blue galaxies. The presence of a UV field affects galaxy formation 
only at low temperatures, which is the case at the beginning of the process, when 
also heating by photo--ionization is important. Moreover, at the
typical densities of strongly cooling, galaxy forming gas ($n_H\gsim 10^{-2}$
cm$^{-3}$), the primordial abundance cooling function is only
significantly modified by a UV field of this kind at temperatures
log$T\lsim 4.5$, even at epochs when the UV intensity is largest ($z=2-3$).

The radiative cooling rate of atomic gas closely depends on its chemical 
composition, 
whose evolution is followed in detail in the simulations (Section 2.3);
in general, this dependence acts in such a way that, at a given
density, the cooling efficiency increases for higher metallicity, leading to
accelerated star formation. 
Including effects owing to metal line cooling is thus decisive for modelling 
in a realistic way the transformation of baryons into stars, and hence the 
whole history of galaxy formation (see Scannapieco {\it et al.}, 2005).

We take metal--dependent cooling consistently into account in the simulations,
by means of the cooling functions of Sutherland \& Dopita (1993: SD): in such
a way the cooling process of the gas takes place accordingly to its metal
content, as resulted from the enrichment described in the 
previous subsection.
With respect to a primordial composition of only H and He, 
metals in the gas affect its cooling rate mainly at temperatures $log T\gsim 4.5$: 
since the effects of the adopted UV field and of metal cooling are relevant
in different temperature ranges (below and above $log T \simeq 4.5$, 
respectively), to a first approximation they can be treated as
independent corrections to the primordial abundance cooling function.

Radiative cooling is a function of the local metal abundance, calculated 
by the standard smoothing kernel of our code: though the metals are not
spread around by diffusion, the cooling function is computed basing on the
average metallicity of all neighbouring gas particles, also in order to
avoid overestimating cooling from particles with high $Z$.

 The adopted metal--dependent cooling functions of SD
 are given for chemical compositions mimicking those in the Solar Vicinity 
 (from solar--scaled to $\alpha$--enhanced at low metallicities).
 In general, the mixture of metals in the simulations (as well as in real 
 astrophysical objects) does not necessarily follow the same trends
 as the local one. Ideally, one would need cooling functions for different
 compositions (different relative proportions of different elements) since 
 the cooling contribution from different elements, different ionization stages 
 and different lines, depends on temperature and a global scaling with 
 metallicity is never a perfect description.
However, this is not yet possible so we link the available cooling
functions to the actual composition of our gas particles, basing on the global
abundance of metals {\it by number}; in fact, what matters for the cooling
efficiency is the number of coolants (metal atoms and ions), rather than the
abundance of metals by mass (the ``metallicity'' $Z$): 
see e.g.\ Section 5.2 of SD.

\subsection{Thermal conduction}

Thermal conduction may play an important role 
in clusters: especially in the richest systems, the highly ionized hot
plasma of the ICM would make up an efficient medium to transport
thermal energy through. Cooling losses at clusters' centre could then get
offset by the heat flow from hotter regions, providing thence a possible
explanation of the apparent absence of strong gas cool-out in observed
clusters at present, which clearly calls for the presence of some additional
heating.

Indeed, simple hydrostatic models assuming cooling and conductive heating
in local equilibrium (Narayan \& Medvedev 2001, Voigt \& Fabian 2004),
have succeeded in reproducing the central temperature profiles, nearly
isothermal with a smooth core decline (see next Section). On the other
hand,
the presence of magnetic fields would be likely to alter the effective
conductivity 
in
clusters, depending on the field configuration itself, which is a still
widely debated topic (Carilli \& Taylor, 2002). Observations of sharp
temperature gradients along cold fronts (Markevitch \etal ~2000, Ettori 
\& Fabian 2000) gives an
indication that ordered magnetic fields are in fact present,
strongly suppressing thermal conduction. 

Thermal conduction was implemented in the code following Cleary \&
Monaghan (1990), with the addition that effects of saturation in
low-density gas (Cowie \& McKee 1977) were taken into account.
In all the runs including thermal conduction, 
we assumed a conductivity of 1/3 of the
Spitzer value (e.g., Jubelgas, Springel \& Dolag 2004), given the
expression
for the electronic heat conductivity in an ionised plasma:
\begin{equation}
\kappa_{Sp}=1.31 n_e \lambda_e k
\left(\frac{kT_e}{m_e}\right)^{\frac{1}{2}}
\propto (kT)^{\frac{5}{2}}
\end{equation}
where $n_e$ and $\lambda_e$ are the electron density and mean free path,
depending on the electron temperature $T_e$ and on the Coulomb logarithm
$\Lambda$.

\section{The gas distribution}
\subsection{Gas density profiles}

The observed gas density in clusters presents a cored profile (usually
parameterized with the ``Beta model" of Cavaliere \& Fusco Femiano, 1976) 
which simulations of
non--merging clusters are not able to explain, due to the underlying cuspy
DM distribution (Metzler \& Evrard 1994).

The gas density profiles of clusters at different temperatures represent
one first clue at their lack or breaking of self--similar scaling, colder
systems having shallower internal profiles. Previous simulations 
(Borgani \etal 2002) 
have confirmed that the scaled density profiles appear identical as
long as only gravitational heating is acting, whereas introducing SN 
feedback causes
the core density to fall in those colder systems where the amount of energy
injected per particle is comparable to the gravitational heat budget previously
available; such effect is even more pronounced when pre--heating the gas by imposing
an early entropy floor. 

\begin{figure}
\includegraphics[width=82mm]{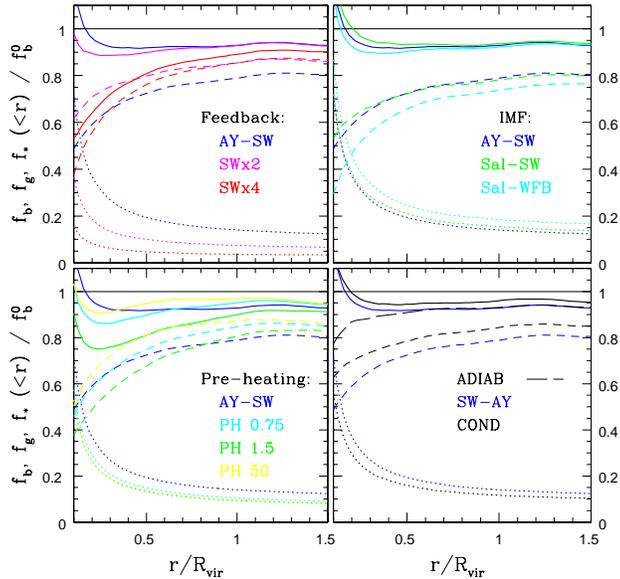}
  \caption{Cumulative mass fractions of the baryon ({\it solid lines}), 
gas ({\it dashed}) and stellar 
({\it dotted}) components of ``Virgo'' cluster, normalized to the baryon 
fraction of the cosmic background used (0.12). In the fourth panel the 
adiabatic curve gives $f_b=f_g$.}                  
\label{fr}
\end{figure}

Directly related to the gas profile is the cumulative gas fraction
within a given radius, namely  
$f_{gas}(r)=\frac{M_{gas}}{M_{tot}}\vert_r$ 
and more indirectly the baryon fraction $f_b(r)$
and the stellar and cold fractions which will be further discussed in Section 7.
In Fig.~\ref{fr} the cumulative mass fractions of the gas and stellar components,
along with the total baryonic one, are shown for the ``Virgo" cluster.
It is commonly accepted (Eke \etal 1998,
Frenk \etal 1999) that simulated clusters without radiative cooling present
gas (and baryonic) fraction profiles converging to a value close to the cosmic value 
at virial
distances, $f_{gas}\rightarrow \frac{\Omega_b}{\Omega_0} = f_b^0$, 
which is eventually reached 
at around twice the virial radius after a gradual rise. 

As can be seen
in the fourth panel, relatively to the
adiabatic run (where $f_{gas}=f_b$, being $f_*=0$), the combined effect of cooling 
and feedback makes $f_{gas}$ to be shifted down by almost 20\% from the core radius 
outwards ({\it cfr.}
with the 25\% value found by Kay \etal 2004), while the
total baryonic profile is unaffected except at the centre, where cooling and 
star formation are responsible for 
the central rise with respect to the adiabatic profile.  
This is in agreement with
other simulations (Kay \etal 2004) pointing out that clusters simulated with 
stellar feedback
exhibit lower gas density than non-radiative ones at overall level, 
the effect being prominent at the centre.
In fact, while cooling removes low-entropy gas from the ICM, feedback 
itself re-heats the gas up:
both processes occurr at a rate proportional to density, therefore reaching 
maximum efficiency near the centre.

The dependence of the radial baryon and gas fraction on the different feedback and/or
pre--heating and IMF recipes can be more specifically examined. The choice of
IMF in practice does not affect neither the baryon nor the gas fractions,
whereas the feedback 
strength can alter the profile: passing from weak feedback to ``standard" SW and 
to SWx2, the baryonic radial distribution remains almost unaffected,
while less gas is turned into stars and the proportion of the hot vs.\ cold 
component changes, as can be seen by inspection of 
the respective $f_{gas}$ and $f_*$  
(Fig.~\ref{fr}: two top panels). This means that (as expected) the stronger the super-winds,
the more gas is left, 
which increases the cumulative (relative) gas fraction from
0.8 to 0.9 
at virial radius: 
in fact, more gas is spread out at larger distances, its profile being
more extended because of higher entropy --- while the star fraction gets 
conversely reduced overall. When moving to 
the extreme SWx4 model, the stellar content is so suppressed at the centre
and the gas distribution so spread out, that the baryonic profile is also significantly 
affected and $f_b$ drops considerably below 0.8 within half the virial radius,
getting to its full (almost cosmic) value well beyond the virial radius.

When adding extra pre--heating to the standard super--wind models (third panel),
it can be noticed that its main effect is the creation of a core in the gas
distribution, which is the more extended, the larger the early energy/entropy 
amount injected per particle.
On the whole, $f_b(r)$ appears to be the most influenced by additional
pre--heating, whereas feedback mainly affects the gas distribution
(unless very extreme feedback efficiencies are adopted).
Both the gas and total baryon fractions increase from the centre outwards,
confirming that the gas distribution is flatter than that
of dark matter (David \etal 1995). 

\begin{figure}
\includegraphics[width=60mm]{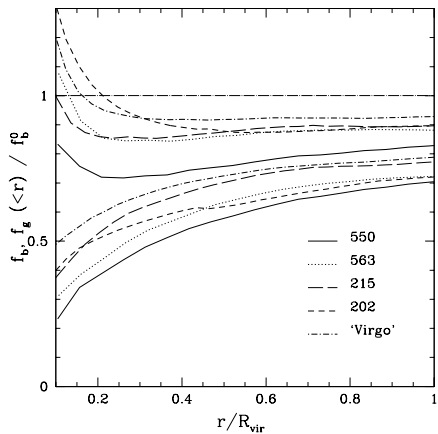}
\caption{Baryon ({\it thick lines}) and gas ({\it thin lines}) cumulative mass fractions of
groups and  ``Virgo'' simulated with the AY-SW standard scheme, normalized as in previous Fig.}
\label{Fall}
\end{figure}

Moreover, when considering the gas and baryonic profiles in simulated
clusters of different mass and temperature (but with otherwise the same 
physical input for feedback etc.; Fig.~\ref{Fall}), the shape of the 
profiles appear rather universal but the absolute density levels scale 
with temperature, with lower central densities and more extended 
profiles at lower temperatures; this is consistent with the findings of 
Vikhlinin \etal (1999) and Roussel \etal (2000).

\begin{figure}
\includegraphics[width=75mm]{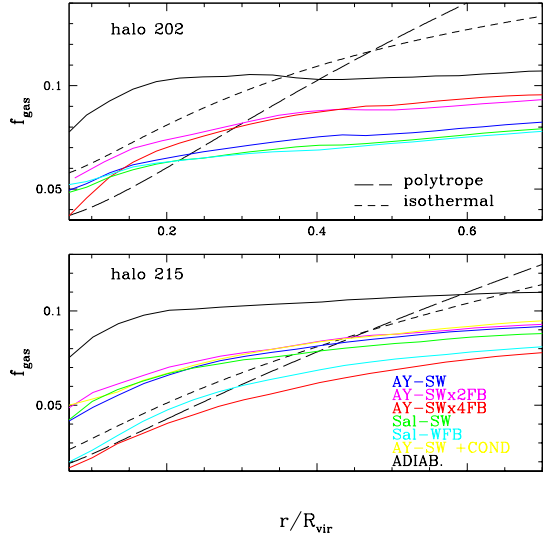}
  \caption{Mass fraction of the ICM gas in the large groups 215 and 202 (all 
models, {\it solid lines}),
compared with those derived by Rasmussen \& Ponman (2004), assuming either a 
polytropic or an isothermal temperature profile on two X-ray bright groups 
of comparable size as our simulated ones ({\it dashed lines}): WARPJ0943.5+1640
and WARPJ0943.7+1644.}                  
\label{fgr}
\end{figure}

In particular it is meaningful to also check for deviations from self--similar
behaviour in the gas mass fraction for lower-mass systems: in Fig.~\ref{fgr} we
present our two small clusters 202 and 215 along with two X-ray bright systems of 
comparable size, for which Rasmussen \& Ponman (2004) have estimated
surface brightness profiles and hence mass and entropy profiles assuming either
an isothermal or a polytropic temperature radial function (see {\it e.g.}
Fig.8 in Section 5); they
found that in both models the gas mass fraction increases with radius, as
expected from a more extended distribution of gas than that of total mass.
But it is interesting to mark as well the relative influence of the cooling
and heating models when applied to different mass systems: while halo 202
still retaines much of the ``Virgo" features, the profile of the smaller 215
indicates that stronger feedback has a relatively large effect on
the ICM distribution at outer distances (see next subsection),
suggesting a $\sim 1.5$ keV scale at which not only gas depletion at 
the centre but even expulsion from the outskirts may become relevant 
if stellar + AGN feedback had a strength as in SWx4 case.

\subsection{Correlations with cluster temperatures}

\begin{figure}
\includegraphics[width=75mm]{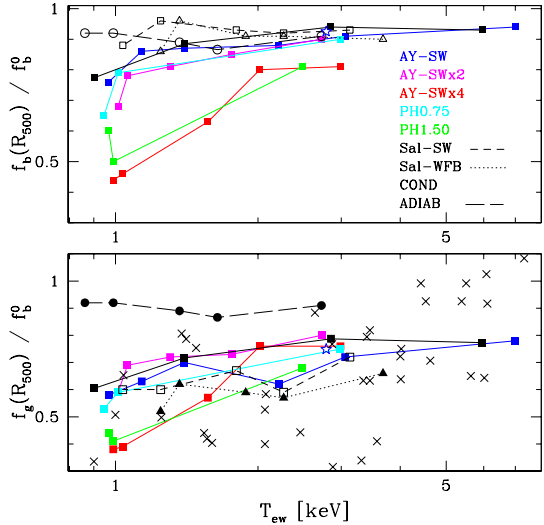}
\caption{Baryon ({\it upper}) and gas ({\it lower})
cumulative gas fractions ($f_b=f_g+f_*$) at $R_{500}$, normalized to cosmic
$f_b$ adopted. Gas data ({\it crosses}) from Sanderson \etal (2003). }
\label{FT2}
\end{figure}

The total baryon fraction distribution is the combined result of the
relative contributions given by its gas and stellar components. Evidence for trends 
in the mean gas fraction with cluster temperature is
ambiguous: a decrease in low--T systems would represent a significant 
contribution to the
steepening of the $L_X-T$ relation (see Section 6), which could be explained by both
lower gas concentration and lower gas fraction (Arnaud \& Evrard 1999), 
especially in models with galactic winds. 
Several studies have found that the stellar component is
predominant in lower-temperature systems, whose lower relative gas content
and inflated gas distribution might be
consequence of a stronger response to feedback in shallower potential wells
(Metzler \& Evrard 1994), although probably not at virial sceles:
indeed this sequence appears very weak even at $R_{200}$, being 
more evident only at inner $R_{2000}$, as found by Roussel \etal (2000).
Sanderson \etal (2003) have however found a ``clear trend" for cooler ($\sim$ 1 keV)
virialized systems to have a smaller mass fraction of emitting gas already at
$0.3 R_{200}$, which gets slightly levelled off when estimated at $R_{200}$
and which is anyway subject to somewhat considerable scatter.
In Fig.~\ref{FT2} we actually find a trend in the gas fraction measured at
$R_{500}$ with temperature, for the same feedback parameter:
in fact the adiabatic
curve lies well above the others (because of no star formation)
and does not exhibit any systematic trend,
whereas an increase is evident in all the AY-SW runs, as well as SWx2 and SWx4 ones.
With this regard, the trend
we predict is in excellent agreement with the data, 
although the large scatter of the latter does not allow us
to single out a best model among the different physical prescriptions
for IMF and feedback.

A constant baryon fraction with size 
would be consistent with the similarity of $f_b$ profiles and an
equally constant gas-to-stellar mass ratio (linked to the cold fraction: see 
Section 7), 
as found by Roussel \etal (2000), indicating that non-gravitational
processes are not dominant in determining the properties of the ICM on
large scale (see also White \etal 1993). However both an increase of gas fraction (albeit modest) and
$\frac{M_{gas}}{M_*}$ with cluster richness have also been separately reported
(Arnaud \etal 1992, David \etal 1995), so that the issue is still debatable. 
From combining the two panels of Fig.~\ref{FT2} it is clear that a trend, if any,
would be imputable to the gas component only, given that the stellar fraction
follows a fairly constant trend. The $f_b$ behaviour confirms that its value 
is noticeably lowered by galactic feedback by almost
one third on average; but above all shows that it is the feedback which is
responsible for the increase of about twice in $f_g(R_{500})$
going from 1 to 5 keV.

\section{Temperature distribution}

Spatially resolved spectroscopical measurements of the ICM temperature
provide crucial information on its thermodynamical state: through assuming
hydrostatic equilibrium, the total gravitational mass of a cluster is
estimable as a function of gas temperature and density profiles. 
X-ray based determinations of the cluster mass generally assume
isothermality for relaxed (non-merging) systems with symmetric
morphology, like those in our sample. 
While to first approximation the ICM appears then to be isothermal,
notwithstanding many recent
data available from {\it ASCA}, {\it XMM}, {\it BeppoSAX}, Chandra satellites for large
samples of rather hot ($T\gsim 4$ keV) systems (Markevitch 
1998, Allen \etal 2001, De Grandi \& Molendi 2002, Ettori \etal 2002, etc.)
agree on presenting declining temperature profiles in the outer regions at
$r\gsim 0.2-0.3 R_{vir}$, displaying a ``universal" shape featuring as
$T\propto[1+(\frac{r}{R_c})^2]^{-\frac{3}{2}\beta(\gamma-1)}$,
with $\beta \simeq \frac{2}{3}$, $\gamma\simeq 1.2-1.3$. Such a decline implies that
the assumption of hydrostatic equilibrium along with isothermal $\beta$-model
of gas density distribution lead to underestimate the gravitating cluster 
mass enclosed within small
radii, whilst overestimating it at large distances. 
In addition, for most clusters hosting cooling flows 
a fair temperature descent was reported towards the centre, after
a peak at around $0.1 R_{vir}$, which reveals the presence of a cooling 
or cold innermost centre surrounded by a somewhat extended
isothermal core:
this feature seems to be reproduced by adiabatic simulations in the literature
(Borgani \etal 2001, Valdarnini 2003), but it vanishes as soon as cooling is
introduced, giving rise to an increasing temperature profile almost to the 
centre. It
has been proposed (Voigt \etal 2002) that thermal conduction, besides
conventional heating, could help regulating central gas cooling: heat would
be transferred from the outer layers to the centre, allowing diffuse gas to be kept 
at fairly low temperature ($\sim$ 1 keV) without cooling out. However, observations
of cold fronts in many clusters indicate that
thermal conduction is significantly suppressed in the ICM.

\begin{figure}
\includegraphics[width=82mm]{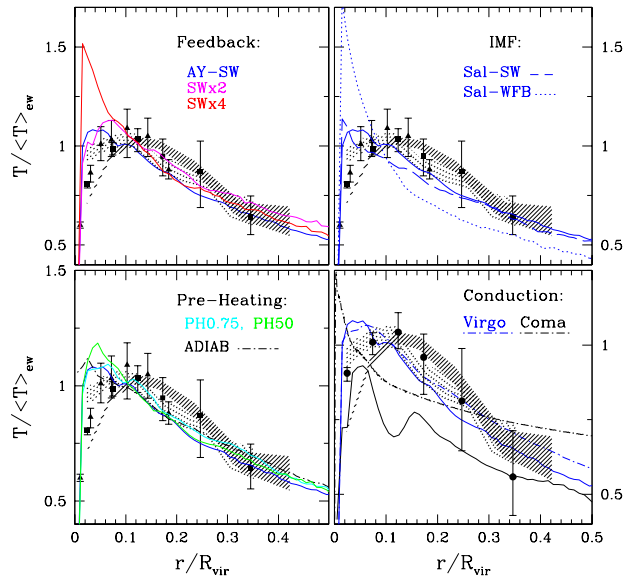}
  \caption{Temperature radial profiles of ``Virgo'', normalized 
to their mean emission-weighted temperature; all panels present AY-SW as reference.
At {\it bottom right} also shown ``Coma''. 
Data points from De Grandi \& Molendi (2002) 
give the average deprojected profile over a sample of nearby clusters, both CF
({\it dash-shadowed})
and non-CF ({\it point-shadowed}); also shown some individual CF clusters: 
A1795 ($<kT>=6.10$ keV, {\it circles}), 2A0335 (3.38 keV, {\it squares}) 
and Centaurus (3.77 keV, {\it triangles}).}                  
\label{TTr}
\end{figure}

In Fig.~\ref{TTr} we compare temperature profiles derived from our
simulated ``Virgo"
with the results by De Grandi \& Molendi (2002), who analysed spatially
resolved data for a sample of rich nearby clusters with {\it BeppoSAX}, both cold--core and 
non cold--core by about the same proportions, finding that their
profiles are characterized by a drop in temperature inside of 0.1$R_{vir}$:
beyond this they steeply decline outwards following a power-law, which is flatter
for the cold--core systems; more specifically, the polytropic indices of
the power-law decline are closer to the isothermal value $\gamma=1$ for
cold--core clusters and to the adiabatic value $\gamma=\frac{5}{3}$ in
clusters without a cold core: the latter are in fact supposed to have 
undergone more
recent major mergers, so that they have not had time enough for heat transport
to efficiently flatten their temperature profiles. 
Such results for cold--core
clusters are also confirmed by Piffaretti {\it et al.} (2005), who give
temperature and entropy profiles of 13 nearby clusters with cooling flows.

In the four panels the separate role of feedback,
IMF, pre--heating and conduction may be assessed by varying the relevant
parameters and keeping the SW scheme with AY IMF as standard reference for all.
The most striking pattern is given by the two extreme runs SWx4 and the
weak feedback one, which are the only ones resulting in an unrealistic 
temperature spike at
the centre. As one can notice in Fig. \ref{TSr}, these are also
the (only) runs displaying a significant high level (``floor") of central entropy, almost
an order of magnitude higher with respect to others, meaning that they must have
a central gas density almost 20 times lower, since $S \propto T/n_e^{2/3}$.
As a consequence, at the same radial distance their gas has a
long cooling time, preventing the onset of cooling flows. 
The reasons for this peculiar central behaviour root
in the structural preparation of these simulations:
the WFB simulation, in which 
gas gets rapidly cooler out of $\sim 0.1 R_{vir}$, adopts an almost 
primordial cooling function due to the very low metallicity 
in the ICM, so that cooling is far less efficient than for the other runs
even in systems larger than groups;
on the other hand the weak propagation
of stellar feedback from the inner sites of star formation causes the gas
to get settled at the centre.
On the contary, the strongest super--wind
scheme ends up with more violently spreading the gas out to larger radii and
thus rarefying the inner density. 
 
With the exception of the Sal-WFB case, all the runs fairly well reproduce 
the observed decline
from $\sim 0.25 R_{vir}$ outwards.
The introduction of strong feedback and/or preheating smooths out the sharp 
central temperature peak of the WFB case (typical also of simulations with
only cooling and star formation) so that the profiles remind more closely
the observational shape;
however, the formation of an isothermal core around 
$0.1 R_{vir}$ is not satisfactorily modelled by any of the simulations: 
temperatures keep rising inwards even beyond $0.1 R_{vir}$ and 
the turn--over occurs only at around $0.05 R_{vir}$.
Further increasing the
feedback strength does not cure the problem, as clearly indicated by the
behaviour of the SWx4 run discussed above (first panel).

In the third panel, preheated runs are compared with the standard AY-SW
and with the adiabatic one: the effect of early injecting 0.75/1.50 keV or
50 keV$\cdot$cm$^2$ per particle are practically equivalent. 
Both the standard and the preheated runs are clearly less flat than the
gravitational-only one at outer distances from $\simeq 0.25 R_{vir}$, while
the latter is steeper at the core scale.

As for thermal conduction, its main effect is to flatten the profile at
large radii, showing that for conductivities of 1/3 the Spitzer value,
this mechanism can even temperature gradients out on large scales.
In particular, for ``Virgo'' conduction does
somewhat smooth the temperature profile around 0.1 $R_{vir}$, 
but by far not enough to reconcile it with the observational profiles.
The ``Coma" simulation discloses instead an interesting feature:
the erratic behaviour of the temperature curve around the core, which by
direct inspection of the simulated cluster is recognized as due to cold
fronts, is completely smoothed out by conduction. Since cold fronts are
observed in many clusters, this indicates that thermal conduction is highly
suppressed in the ICM, presumably by magnetic fields.

\section{Entropy}
The dynamic and X-ray properties of the ICM of relaxed clusters are
on large scale determined by the entropy distribution of the inter-galactic 
gas adjusting itself to fit into the dark matter potential well it lies within.
The intra-cluster gas entropy (defined as $S=kT/n_e^{2/3}$, a quantity 
conserved in adiabatic processes) gravitationally generated across 
the growth process of cosmic structure changes its distribution by means of 
radiative cooling and non-gravitational heating, and in turn determines gas 
properties such as density and temperature profiles (Voit \etal 2002; Borgani 
\etal 2001).

Observations of X-ray surface brightness profiles in inner regions of
poor clusters and groups ($T\lsim 2$ keV) had pointed towards the existence of an 
``entropy floor'' in the ICM/IGM at around 100 keV$\cdot$cm$^2$ (Ponman \etal, 1999,
Finoguenov \etal 2002),
breaking the self-similarity of group and cluster ICM
profiles. Systems at all scales, and in particular low-mass ones, show 
evidence for
excess entropy compared to that expected from pure gravitational entropy increase
by shocks (see Introduction). 

In our approach we also included a set of simulations with additional 
preheating, basically following Borgani \etal (2001, 2002) in setting an 
entropy floor through early instantaneous preheating of 0.75-1.50 keV or 
50 keV$\cdot$cm$^2$ per particle at $z=3$, which is the epoch at which heating 
sources are expected to contribute significantly (close to peak of star
formation in proto-clusters). An impulsive energy/entropy budget of the
order we used is motivated by X--ray phenomenology of ICM, and appears
also consistent with detected abundances (see Tornatore \etal 2003).

However, observations of groups have recently hinted towards a lack of
large isentropic cores, at variance with predictions of 
preheating models (Ponman \etal 2003). Especially massive virialized systems with
strong cooling flows do not show such cores, rather a monotonically
increasing entropy with radius (Piffaretti {\it et al.} 2005 and references therein). 
On the other hand, high entropy 
excesses have been reported up to large radii even in moderately rich clusters 
(Finoguenov \etal 2002). In fact, whilst the existence of isentropic
cores is still controversial, both 
analytical estimates of the effect of shock heating on the preheated accreting 
gas (Tozzi \& Norman 2001) and numerical simulations (Borgani \etal 2001) 
indicate entropy profiles outside the cluster's central region
having a slope $S\propto r^{1.1}$, or closely shallower. 

Another explanation could appeal to cooling alone as able to remove 
low-entropy gas from the haloes' centre, but 
this class of models inevitably leads to overcooling (Knight \& Ponman 1997, 
Dav\'e \etal 2002).
It is clear that the most natural and comprehensive way to model the inner 
distribution of entropy, as well as the X-ray emission, is to accompany cooling 
with heating by feedback subsequent to star formation: this will heat gas 
particles at low entropy.
The entropy threshold in the core would be then set up by equating the gas 
cooling time to a Hubble time, such that the floor itself will scale as 
$S_f\propto (\tau_{cool}T)^{2/3}$ and will be higher in hotter systems, 
as to reproduce the observed smooth flat trend with temperature
(Voit \& Bryan, 2001).

\begin{figure}
\includegraphics[width=80mm]{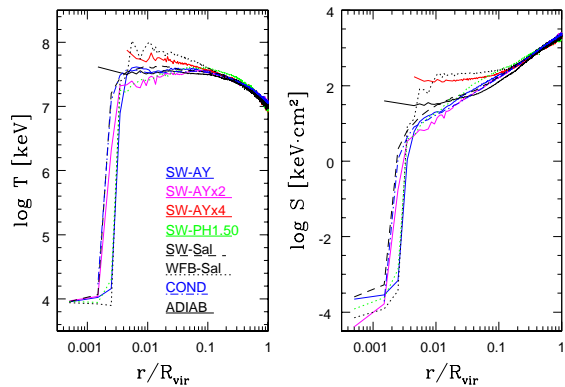}
  \caption{Temperature ({\it left}) and Entropy ({\it right}) profiles of 
``Virgo'', all models.
}  
\label{TSr}               
\end{figure}

In Fig.~\ref{TSr} the inner pattern of temperature and entropy for 
particles in the ``Virgo'' cluster is shown in
more detail, from which it is evident that there is not a proper isentropic 
core or entropy floor: most models exhibit central entropy values of
$\sim 10-20$ keV$\cdot$cm$^2$, in agreement with inferred values
for cold--core clusters (Voit \etal 2003); instead a plateau 
around 100 keV$\cdot$cm$^2$ has formed only 
in two distinct cases, namely the WFB scheme with Salpeter 
IMF and the 4-times SW scheme with AY IMF. 
The former has less efficient cooling 
and therefore does not develop a low-entropy core; at the same time,
the higher star formation rate (because less hampered by weaker feedback)
also contributes to absorb larger amounts of low--entropy gas. 
The SWx4 model 
ends up with forming a floor around 100 keV$\cdot$cm$^2$ at $z=0$,
much higher than the entropy level reached by the preheated runs;
preheating indeed seems to have a minor impact on the behaviour 
of the standard SW simulations. The adiabatic run also develops 
an entropy floor of about 30 keV$\cdot$cm$^2$, but only at the very centre 
($r\leq 0.01 R_{vir}$),
where gas keeps accumulating without cooling.

\begin{figure}
\includegraphics[width=80mm]{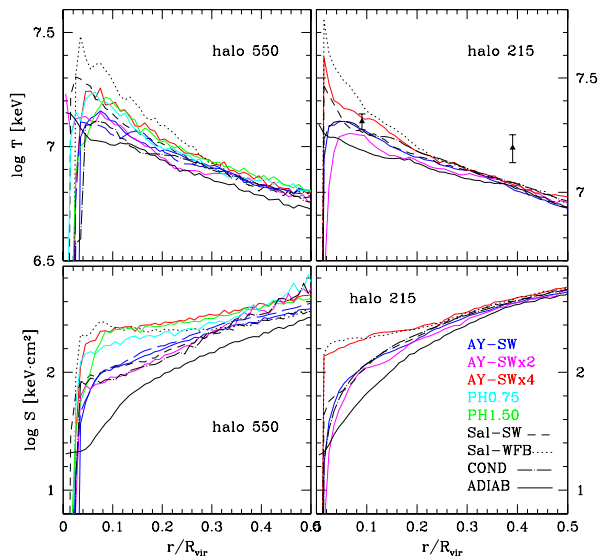}
  \caption{Temperature ({\it top}) and Entropy ({\it bottom}) profiles of the
groups 550 and 215 (all models) at $z$=0. Also shown temperature profile of group
550 at high (8x) resolution (thick {\it long-dashed} lines). 
T-data points from Rasmussen \& Ponman, 2004.}
\label{TSgroups}                  
\end{figure}

The entropy floor is even more evident in our simulations of groups 
(Fig.~\ref{TSgroups}), where not only the aforementioned WFB and SWx4, but also 
the preheated runs, exhibit evidence for an isentropic core at $\sim$ 100-250 
keV$\cdot$cm$^2$ up to $\sim 0.3 R_{vir}$; while the adiabatic one clearly 
lies the lowest (except that at the innermost $\sim 0.05 R_{vir}$) and all 
other profiles fall in between, following a smooth coreless increase from the 
centre. 
In this way, coupling cooling and star formation with feedback 
allows one to 
reproduce the entropy excess observed at least at groups scale: namely, 
cooling and star formation alone make low-entropy gas particles be selectively 
removed from the diffuse hot phase in central regions, hence leaving a 
flatter profile. Imposing then an artificial entropy floor at high redshift 
onto this has a slight effect on increasing the gas cooling time and 
thus slowing down star formation, so that the final entropy level gets lower 
due to less efficient cooling. When properly incorporating 
energy and metal feedback from star formation, one can see 
how the gas entropy actually ends up at a lower level. 

\begin{figure}
\includegraphics[width=60mm]{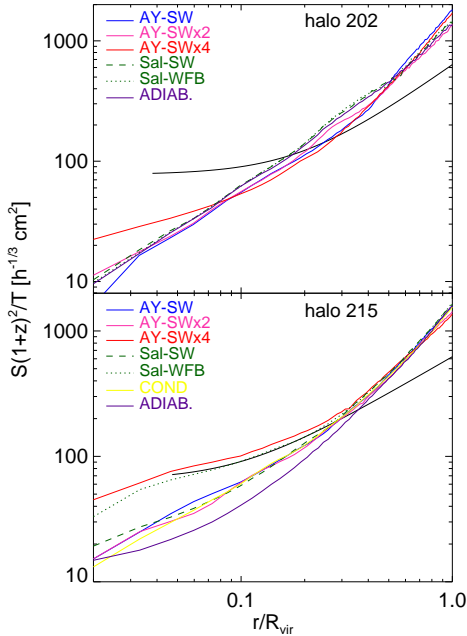}
  \caption{Scaled entropy as a function of radius. {\it Dot-dashed:} normalized 
self-similar profile $S \propto r^{1.1}$ expected from pure gravitational shock-heated 
spherical accretion. {\it Thick lines}: scaled entropy in the two bright groups of 
Rasmussen \& Ponman (2004), obtained by assuming polytropic gas temperature profiles.}
\label{EntrJR} 
\end{figure}

In order to highlight the effects of non-gravitational physics on the gas 
radial distributions, we show in Fig.~\ref{EntrJR} the scaled entropy 
$S/T$ for two groups, whose characteristics are comparable to 
those of the X-ray bright galaxy groups studied by Rasmussen \& Ponman (2004). 
Indeed self-similar scaling with mass would lead to 
the relation $S(r) \propto T$ and spherical shock heating models 
as well as simulations predict the power law $S(r)\propto r^{1.1}$,
which is also plotted for comparison. In general all our simulated profiles
get steeper than this when approaching the virial radius, whilst they flatten 
inside $\sim 0.1 R_{vir}$ as a result of a core in the density distribution. 
The mean slope outwards $0.1 R_{vir}$ appears however to follow that for
clusters (see Piffaretti {\it et al.} 2005).
The only evident 
entropy core has formed again in the SWx4 simulation and in any case no core 
at all can be seen extending beyond this rather small radius: 
this is consistent with some recent observations of groups ({\it e. g.} 
Pratt \& Arnaud, 2003), which do not detect in poor systems the large 
isentropic cores predicted by simple preheating models.

\section{X-ray Luminosity}

The relationship between
bolometric luminosity and temperature provides a sensitive diagnostics of the 
gas to virial mass ratio and an indication as well of the lack of self--similarity in
the structure of clusters. 

Measurements of ICM temperature by X-ray
satellites demonstrate that the observed systems scatter around the curve 
$L_X\propto T^\alpha$ with $\alpha \simeq 3$ for clusters at $T\gsim 2$ keV
(Mushotzky 1984, Edge \& Stewart 1991, David \etal 1993, Markevitch 1998, 
Arnaud \& Evrard 1999, Ettori \etal 2002), approaching to the self-similar
scaling only from $T\gsim 8$ keV on. 
The relation itself has been found fairly constant over
redshift up to $z\sim 1$ (see Donahue \etal 1999, Borgani \etal 2001).
The largest deviations from it
are found for those systems whose emission is strongly associated with a
cold core, or ``cooling--flow" (Fabian 1994), 
whose effect has to be corrected 
for (Markevitch 1998, Allen \& Fabian 1998, Arnaud \& Evrard 1999, 
Ettori \etal 2002). 
This class of clusters represents slightly more than half of all
nearby X-ray detected clusters. 
Moreover there is evidence that low-mass systems 
($T\lsim 1$ keV) show a steeper relation, up to $L_X\propto T^8$ 
(Ponman \etal 1996, Helsdon \& Ponman 2000b). 

As sketched in the Introduction, this discrepancy with gravitational-only 
(non-radiative, or
``adiabatic") hierarchical models of spherical accretion indicates
a segregation of dark and gaseous matter.
Pre--heating of the
ICM later followed by merger shocks could
steepen the $L_X-T$ dependence of cooler clusters (Cavaliere \etal 1997) and at
the same time decrease the luminosity evolution to redshifts $z\sim 1$,
as observed. When cooling and star formation were included in the simulations, the 
resulting emission 
profiles got flatter at the core due to gas removal from the hotter phase, while 
incorporating also additional heating on this has produced spikes
in the inner X-ray luminosity density if the epoch of pre--heating was as
early as $z=9$ (Borgani \etal 2001). Tornatore \etal
(2003) have found that cooling plus star formation only are able to fairly suppress
the X-ray luminosity, but at the same time are responsible for steepening the
temperature profiles, leading to an increase in the emission--weighted temperature; 
the net effect is to enable a match between simulated and observed clusters
in the $L_X-T$ plane, though not at group scales.
Including later ($z=3$) pre--heating mainly affects small systems, further
suppressing their emissivity, so that altogether
the combined effect of cooling and pre--heating is for simulations 
to approach the slope of the observed $L_X-T$ relation.

Besides cooling, star formation and energy feedback, our procedure included as well
a self--consistent treatment of the metal enrichment of the ICM regulated by the
stellar feedback itself, in such a way that it is possible to estimate also
the contribution heavy elements make to the emissivity. 
In general each gas particle has an assigned
mass, density (hence a volume $\Delta V_i$),
temperature, metallicity $Z_i$ and energy, so that its bolometric luminosity 
at X-ray band can be computed as
\begin{equation}
L_X= \sum_i L_X^i = \sum_i \Lambda (T_i)_{Z_i} \cdot n^2_{H_i} \cdot \Delta V_i
\end{equation}
where $\Lambda$ is the metal-dependent cooling function such that the emissivity 
reads
$\epsilon = \Lambda(T,Z)\cdot n_H^2$, and which scales as $\Lambda(T,Z)\propto \sqrt T$ for
$T\gg 10^6$K. The contribution of metal lines to global emissivity, which is
dominated by Bremsstrahlung from a primordial-like gas, gets even more relevant
at group scales, where it can make up to half of the total. From this, the 
emission-weighted temperature is obtained as
\begin{equation}
\langle T \rangle =\frac{\sum_i L_X^i\cdot T_i}{\sum_i L_X^i}
\end{equation}

\begin{figure}
\includegraphics[width=75mm]{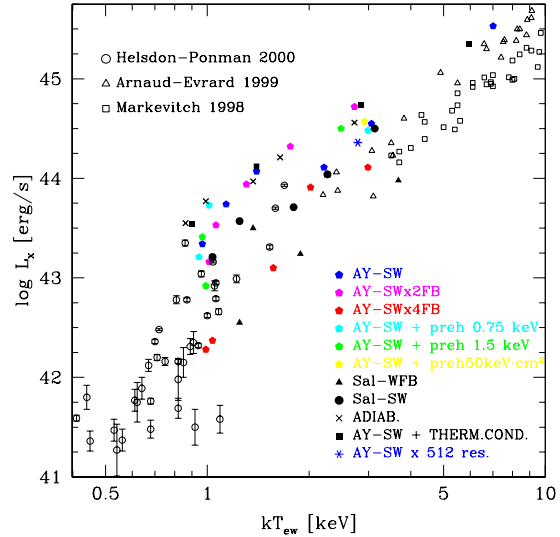}
  \caption{X-ray luminosity - emission weighted temperature scaling 
relation from the inner 1 Mpc. Simulated points are the average over the last 
ten frames down to z=0. High-resolution Virgo shown at $z=0.1$. {\it Open} symbols
are X-ray data from Markevitch (1998), Arnaud \& Evrard (1999), Helsdon \& Ponman (2000a).}
\label{LT}
\end{figure}

In Fig. \ref{LT} the X-ray luminosity--(emission-weighted) temperature relation 
for all our simulations at $z \simeq 0$ is shown,
along with observations. The observed (open) points exhibit the characteristic double-slope
behaviour; as expected, the adiabatic runs lie above this curve, but interestingly 
adding
thermal conduction results in a similar overestimate of the X-ray emission as well.
Overall, the predicted luminosities are higher than the observed ones; both enhanced
feedback and a less top--heavy IMF, however, can 
lower the X-ray luminosity: the latter effect seems to be dominant,
since the runs with Salpeter IMF and Weak Feedback, which result in a low
level of metal enrichment as well (see next sections), are those placed lowest. 
All in all, when strengthening the feedback up to 4 times and using a top--heavy
IMF, a satisfactory level seems to be reached at least for 
clusters, whereas it results in a too drastic
suppression of the X-ray luminosity for the lower mass haloes, 
though matching the slope of the $L_X-T$ relation.
A preliminary conclusion from these results could
be that the success of previous generations of simulations in reproducing both
the L-T relation, the surface brightness profiles and the rather high
entropy floor, may have been somewhat favoured by the simplistic use of primordial or 
uniform cooling functions; when making the models more
realistic by including chemical evolution and metal-dependent radiative cooling, this
promotes the development of fairly strong central cooling flows, which generally lead 
to excess X-ray luminosities.

\begin{figure}
\includegraphics[width=60mm]{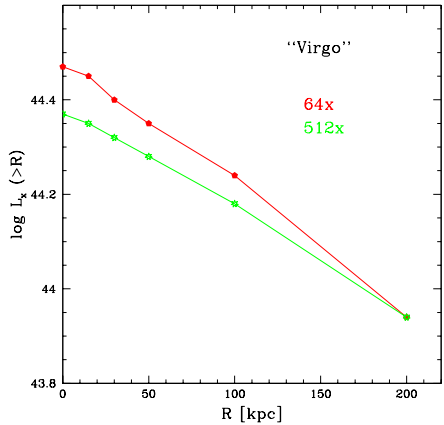}
  \caption{X-ray luminosity emitted from the gas contained between $R$ and 
$R_{vir}$, for ``Virgo'' run at normal ({\it red, pentagons}) and high 
({\it green, stars}) mass resolution, both at $z=0.1$.}
\label{Lr}
\end{figure}

Finally, in Fig.~\ref{Lr}, we show the effect of the numerical resolution
on the computation of the luminosity, in the case of ``Virgo'': as it
can be noticed, increasing mass resolution by 8 times and force resolution 
by twice, affects only slightly the global X-ray luminosity,
the largest contribution to which being that of the innermost 
denser region, 
which is now better resolved.

\subsection{X-ray luminosity and star formation evolution}

Besides the ICM, star forming galaxies too are known sources of X-ray emission,
which can be considered as a tracer of the most energetic phenomena
associated with phases of star formation and evolution, 
such as mass accretion, SN explosions and remnants, hot stellar winds and
galactic outflows. While 
point sources (X-ray binaries, SN remnants) account for the dominant 
contribution to the overall X-ray emission in the 2-10 keV band, where they 
give the spectrum a power--law shape, diffuse emission (the thermal 
component) is significant in the soft X-ray band, arising from the ISM shock 
heated by supernov\ae\ in the form of shells or winds (Persic \& Rephaeli 2002). 
Moreover, AGN X-ray emission correlates with enhanced star formation 
(starburst) sites: the X-ray emissivity normally overwhelms that 
from the host galaxy itself.

A relationship between far-infrared (FIR) and X-ray luminosities in star forming
galaxies has often been found in both soft and harder X-ray bands ({\it
e. g.} David \etal 1992; Lou \& Bian 2005 from ROSAT in the soft 0.1-2.4 keV band;
Ranalli \etal 2003 from {\it ASCA}, {\it Bepposax} and Chandra in the hard band;
Franceschini \etal 2003 from {\it XMM} in the hard 2-10 keV band): since radio 
and FIR luminosities are well known
indicators of star formation, it follows that $L_X$ at both 0.5-2 keV and
2-10 keV should as well provide an estimate of the star formation rate (SFR).
Besides, spiral galaxies in the field or cluster outskirts may still be
embedded in extended reservoirs of hot emitting gas still accreting onto their discs.
The gas accretion history of the disc is thus indirectly reflected by the
galaxy X-ray luminosity: the hot gas infalling from the halo cools out with
a mass cool-out rate inversely proportional to its cooling time and following
Rasmussen \etal (2004) it approximately scales with the
bolometric luminosity and the emission-weighted mean inverse temperature:
$\dot{M} \propto L_X \langle \frac{1}{T}\rangle_{ew}$. 

\begin{figure}
\includegraphics[width=75mm]{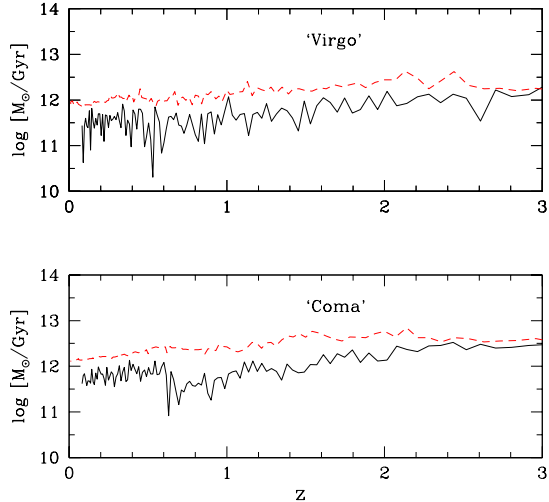}
  \caption{Star formation rate ({\it solid}) in ``Virgo'' (AY-SW) and ``Coma'', 
including the cD 
and all the stars (both inter- and intra-galactic), compared with the 
mass cool-out rate $L_x\langle \frac{1}{T}\rangle_{ew}$ by X-ray emission 
({\it dashed})
as predicted by simple model in Rasmussen {\it et al.}, 2004.}
\label{evol}
\end{figure}

In Fig. \ref{evol} we check for the same
relationship in tracing the redshift evolution of the predicted gas cool-out
rate and global star formation rate of
the whole cluster for the two largest simulated systems: it is noticeable
that the two quantities are well correlated, although the predicted cool-out rate is
systematically higher by a factor $\sim$3. 
The offset between the two quantities is due to: a) the
simplistic assumption behind the equation we took from Rasmussen \etal
(2004), where $PdV$ work was neglected; and b) the fact that, given the strong galactic
super-winds, not all gas which cools out is turned into stars.

\section{The cluster cold components}

The baryonic mass of clusters is largely dominated by the hot ICM mass,
yet the mass fraction of the cold component (which is the sum of stars
and, although to a much smaller extent, the neutral, non X-ray emitting 
gas phase at $T\simlt 10^4 K$)
is a crucial constraint for cluster simulations, as it plays a major role 
in distinguishing the mechanisms responsible for the ``excess entropy'': 
the larger the cold fraction, the larger the contribution given by 
star formation in removing low entropy gas from the ICM.
To match the observed cold fraction of $f_c=0.1-0.2$ (see below), 
strong (pre)heating or feedback are needed to enhance the entropy level 
(see Section~3.4). Also, the cold fraction constrains the chemical enrichment
efficiency of the stellar populations --- which are evidently able to enrich 
to significant levels the hot ICM gas mass $M_{ICM} >> M_*$ (Section~8).
The cold fraction, defined as the fraction of baryons in the form
of cold gas and stars, provides a measure of galaxy formation efficiency
and can be recast in terms of the gas and stellar mass-to-light ratios (MLR) as
\begin{equation}
f_c^{-1} \simeq
1+\frac{M_{ICM}}{L_B}\left(\frac{M_*}{L_B}\right)^{-1}
\end{equation}

Its empirical estimate however is non--trivial:
while the ICM gas mass is determined fairly directly from X-ray data, 
the stellar mass $M_*$ can be derived only indirectly from the observed 
luminosity of cluster galaxies, typically in the B band, and is subject 
to uncertain assumptions on the stellar mass-to-light ratio; the latter likely
spans the range 4-8 $\frac{M_\odot}{L_\odot}$, according to the IMF
(Portinari {\it et al.} 2004). 
Thus, with the ratio $\frac{M_{ICM}}{L_B}$  directly observable and 
typically $\simeq 30 h^{-\frac{1}{2}} \frac{M_\odot}{L_\odot}$ 
(Moretti {\it et~al.}\ 2003; Finoguenov {\it et~al.}\ 2003)
\footnote{As the 
ICM gas MLR tends to increases with the sampled
volume, both in simulations (where the gas to star ratio increases outward,
Fig.~1) and in observations (Roussel {\it et~al.}\ 2000), it is important 
that comparison to observational data is performed within the same radius. 
In the following, we will discuss the cold fraction within $R_{500}$, 
unless otherwise stated.},
the corresponding cold fraction $f_c \simeq 10-20 \%$ has a factor-of-two 
uncertainty stemming from the assumed stellar MLR (see Portinari 2005 
for a detailed discussion). Due to the importance of this constraint
on cluster physics, care should be taken in comparing simulations
and observational estimates consistently (i.e.\ assuming a coherent stellar 
MLR), especially since the Salpeter and AY IMFs of our simulations 
are quite ``heavy'', i.e.\ with a significant stellar mass locked 
in low--mass stars and remnants (Portinari 2005). 
The corrections can be sometimes quite significant (e.g.\ in Fig.~\ref{FcR}).

\begin{figure}
\includegraphics[width=75mm]{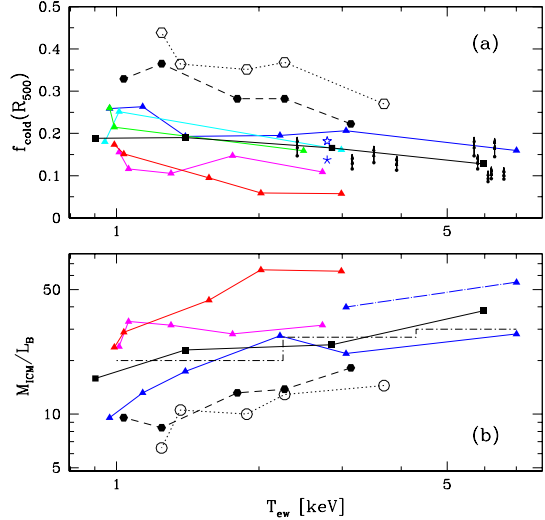}
  \caption{(a) Cold fraction at $R_{500}$ and $z$=0. 
``Virgo'' is shown at both normal and high 
resolution: the latter (at $z=0.1$) with ({\it star}, since $z=2$) 
and without ({\it open star}) cooling flow correction. 
Data in the {\it K-}band from Lin
\etal 2003 are provided with an upward arrow indicating a 30\% (20\%)
shift to consistently match the simulations with Salpeter (AY) IMF (see text).
(b) Mass-to-light ratio of the hot ICM gas in the {\it B-}band. 
Luminosities have been calculated from all the stars 
formed within $R_{500}$: for ``Virgo'' and ``Coma" is also shown ({\it dot-longdashed})
the effect of dropping the young stars at the very centre of the cD (see text).
Data from Finoguenov {\it et al.} 2003 ({\it dot-dashed}: average over cluster
samples).
Colours as in Fig. 4 and 18. } 
\label{CFT}
\end{figure}

In Fig.~\ref{CFT}a, the cold fractions of our simulations
are compared to the recent data by Lin {\it et~al.}\ (2003; hereinafter LMS) 
who accurately derive stellar masses from K--band luminosity (a better mass 
tracer than B band), distinguishing between spirals and ellipticals 
in assigning the MLR. Yet the zero--point of the {\it absolute value} 
for the stellar MLR depends on the assumed IMF;
the MLRs adopted by LMS (from Bell \& de Jong 2001 for spirals and Gerhard 
{\it et~al.}\ 2001 for ellipticals) are about 30 (20)\% smaller than those 
of the Salpeter (AY) IMF. 
This implies that, by assuming a Salpeter (AY) IMF consistent with 
our simulations, the underlying stellar mass and the cold 
fraction would be 30 (20)\% larger than inferred by LMS; these corrections are
shown as vertical shifts in Fig.~\ref{CFT}, nonetheless they are by far not 
yet sufficient 
to bring the Salpeter simulations in agreement 
with the data; especially those with weak feedback predict too large $f_c$.

Good agreement with the data is obtained instead
for the ``standard'' AY-SW simulations, with or without pre-heating and 
conduction.
In Fig.~12a we show the cold fraction of the
high resolution ``Virgo" simulation (quite close to
the normal resolution one), along with the one
corrected for the mass of the central cooling
flow stars (Section 2.2): neglecting the latter population reduces 
$f_c$ by about 20\%. 
As to the simulations with enhanced feedback efficiency,
the SWx2 simulations are still in reasonable agreement with the
observations, while the extreme case SWx4 yields too low cold fractions
for $T>2$keV, where data are available.

Alternatively, rather than correcting the observational cold fraction 
consistently with the specific IMF and MLR pertaining to each simulation, 
we can compute the stellar luminosity of our clusters and
compare to the directly observed $\frac{M_{ICM}}{L}$ ratio.
The global luminosity within $R_{500}$ is the sum 
of the luminosities of all the star particles, each computed from SSPs
with the corresponding age, metallicity and IMF (see Paper~II for details).
Again, when computing the cluster luminosity, we neglect the artificial late stellar
populations arisen from innermost cooled region (see Section 2.2), in the same 
way as for the ``Virgo" cold fraction above.

In Fig.~\ref{CFT}b we plot the ICM MLR from our simulations, compared to the
{\it B}-band data of Finoguenov {\it et al.} (2003),
for volumes within 0.4~$R_{vir}$.
In agreement with the top panel, the AY-SW simulations best reproduce
$\frac{M_{ICM}}{L_B}$ and hence the cold fractions; also the trend with 
temperature is well modelled.
For the extreme feedback case AY-SWx4, the low cold fractions
on top corresponds as expected to high $\frac{M_{ICM}}{L_B}$.
For the Salpeter simulations, also consistently with the too large cold fractions 
found above, the luminosity is too large for the corresponding ICM mass and
$\frac{M_{ICM}}{L_B}$ is too low; the excess luminosity is even more
striking in B band, because in the Salpeter simulations the
galaxies are more metal poor and hence bluer, than real cluster galaxies 
(see the colour-magnitude relation in Paper~II). 

A number of studies suggest that the cold fraction decreases at increasing
cluster mass or X-ray temperature (David {\it et~al.}\ 1990; 
Arnaud {\it et~al.}\ 1992; LMS) --- though the issue is still debated, 
and it might depend on the sampled volume in different clusters 
(Roussel {\it et~al.}\ 2000). 
The trend is usually explained by invoking less efficient star and galaxy 
formation in hotter systems; indeed LMS have recently found that the stellar 
mass fraction over the total cluster mass, decreases for more massive 
clusters. 
To this regard, an important role is also played by metal dependent 
cooling, especially when coupled with efficient mechanisms of transportation
and mixing of the heavy elements into diffuse gas at outer radii, which has yet
to cool out (Scannapieco \etal., 2005): thus, in more massive haloes the 
effects from non-primordial
cooling function (as that we use, except in the WFB model) are possibly
counteracted by chemical mixing occurring at larger scale.
Another factor contributing to the trend is the likely gas dispersal out of
the shallower potential wells of smaller systems, reducing their
ICM mass and overall baryon fraction.

For our simulated sample as well, the cold fraction decreases with increasing
mass (Fig.~\ref{CFT}a). The trend is driven by the fact that
larger clusters better retain the ICM gas, so that the ICM to total mass 
ratio increases,
while the stellar mass fraction remains fairly constant with cluster 
temperature (cf.\ Fig.~\ref{Fall} and~\ref{FT2}). 
The latter behaviour is in conflict with observations;
notice, however, that the observed stellar mass fraction typically refers
to the stars in galaxies: the intra-cluster population is not taken
into account here and it might compensate for the observed descrease, 
if the percentage of intra-cluster stars increases with cluster mass, 
due to more efficient stripping (LMS; see Paper III).

\begin{figure}
\includegraphics[width=70mm]{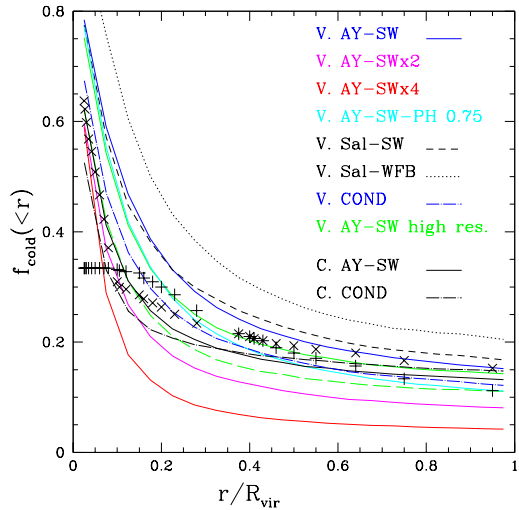}
  \caption{Cold gas plus stellar radial profiles of the two clusters. 
Also shown ``Virgo'' at higher resolution ({\it thin} lines), both with ({\it
long-dashed}) and without ({\it solid}) cooling flow correction since $z$=2. 
The observed cold fractions derived from $\frac{M_*}{M_{ICM}}$ of Roussel 
{\it et al.} (2000) are 
overplotted (+ for King light profiles, $\times$ for de Vaucouleur's).}
\label{FcR}
\end{figure}

Non-gravitational heating mechanisms (SN/AGN-driven galactic winds; 
pre-heating) also affect the radial profiles of the cold fractions.
In Fig.~\ref{FcR} we compare our two largest simulated clusters to the 
observed profiles of Roussel {\it et al.} (2000)\footnote{The profiles 
of Roussel {\it et al.} have been scaled to our adopted value 
of $H_0$ (since $L_B/M_{ICM} \propto h^{\frac{1}{2}}$), 
and from their assumed, very low stellar MLR=3.2 to one consistent with 
the AY IMF. Altogether, the data in Fig.~\ref{FcR} have been scaled by 
a factor of 2.};
the + and $\times$ symbols correspond to the results they obtain fitting King
or de Vaucouleur light profiles, respectively; the latter are
better comparable to our simulated clusters where the light is centrally 
peaked due to the cD's. 
Once more the AY-SW simulations (including the conduction
and preheating cases) are those that best reproduce the overall cold fraction
within $R_{500}$.
Again, as previously in Fig. 12, the effect on the cold fraction of removing
stars belonging to the base of the cooling flow is highlighted:
in particular the high resolution and corrected 
``Virgo" simulation (dashed green line) turns out to be in very close agreement 
with observed profiles.

All in all, it is a positive result that our simulations with supernova feedback 
(with a top-heavy IMF, but without other energy sources such as AGNs)
fairly reproduce the observed low cold fractions.
Previous simulations including radiative cooling 
and star formation typically predicted too large a fraction of gas 
converted into stars, with SPH as well as with Adaptive Mesh Refinement 
schemes (Tornatore \etal 2003 and references therein; Kravtsov,
Nagai \& Vikhlinin 2005). When artificial pre-heating is also invoked,
star formation is hampered at high redshift, which helps to reduce the cold 
fraction but also causes an unrealistic delay in the star formation history
(Paper~II; Tornatore {\it et~al.}\ 2003; Nagai \& Kravtsov 2004). 
Dumping the supernova energy just as thermal energy in the surrounding
gas has little effect on the cold fraction (Tornatore {\it
et~al.}\ 2003; Nagai \& Kravtsov 2004); hence the key to our result is 
a feedback implementation able to represent more realistically the
adiabatic phase of super-shell expansion.


\section{Metal enrichment}

The hot ICM gas in clusters is significantly enriched in metals.
In our simulations, the super--winds resulting from SNII 
enrich the ICM by dispersal of the chemical elements 
produced by stars (Section~2.3 and ~2.1). In this scheme, the dispersal 
of SNII products is favoured; however, metals ejected by lower mass stars 
and SNIa are blown out 
as well (and are important for the ICM enrichment), because the super-winds 
also expel gas already present in the galaxies.

As to the role of ram pressure stripping, though in principle included 
in the hydrodynamical simulations, it is here only partially resolved:
stripping of hot and diffuse galaxy halo gas is resolved, whilst stripping
of high density, cold (HI) gas in the galaxy itself is not.
There are however arguments that ram pressure stripping
is not the dominant mechanism of extraction of metals from galaxies (Renzini 
1997; Aguirre {\it et~al.}\ 2001, but see also Domainko {\it et al.} 2006 and
Section 8.4).

\begin{figure}
\includegraphics[width=70mm]{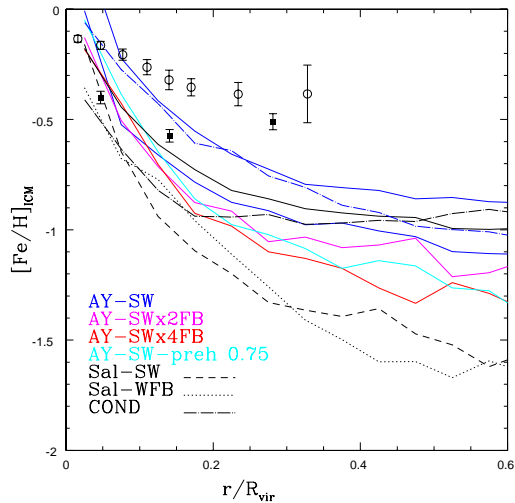}
  \caption{Iron gradients at $z$=0 in the ICM of the two simulated
clusters (same symbols as previous figure: ``Virgo'' at higher resolution as
thinner solid line). Data from De Grandi et al. (2004) refer to coldcore 
({\it open circles}) and non-cold core ({\it filled squares}) clusters.}
\label{FehVC}
\end{figure}

\subsection{The iron content of the simulated clusters}

In rich clusters the iron abundance is about one third of the solar value, 
with a rather small dispersion (Arnaud {\it et~al.}\ 1992;
Renzini 1997; Fukazawa {\it et~al.}\ 1998; Baumgartner {\it et~al.}\ 2005);
detailed iron profiles suggest though that cool core
(formerly called cooling flow) clusters have systematically higher ICM 
metallicities (De Grandi {\it et~al.}\ 2004, hereinafter DELM; 
Fig.~\ref{FehVC}).
The iron abundance profiles show negative gradients: 
cool--core clusters display a central peak (up to 0.7 solar), 
while out of the cool core, as well as in non--cool core clusters, 
the gradients are much milder and consistent with a flat distribution
(DELM). 

Fig.~\ref{FehVC} compares the iron abundance profiles in the ICM from our 
simulations, to the data for cool core and non-cool core clusters; the former
are a better reference for our simulated clusters,
which are well relaxed objects with 
prominent central cDs. Evidently, our profiles
are steeper than the observed ones --- a general
problem in simulations (Tornatore {\it et~al.}\ 2004), 
partly related to the fact that star formation at late times is concentrated 
toward the central regions.

In Fig.~\ref{FehVC}, the iron abundances of the Salpeter 
simulations lie well below the observed ones, at all radii; while the
standard AY-SW simulations are closer to the data. However, since the 
distribution of the metals is different between simulations and real
clusters, quantitative comparison of the level of chemical enrichment 
can be made considering the {\it global} metal content of the ICM 
only.\footnote{Rather than the global iron content (mass weighted metallicity) 
of the ICM, we could discuss the emission-weighted metallicity, which 
in principle corresponds to the direct observational quantity. 
We prefer not to follow this line, since in the
simulations the very steep gradients artificially boost the emission weighted 
metallicity, biased in favour of the central, brightest regions, by a factor 
of 2--3 (e.g.\ Tornatore {\it et~al.}\ 2004). Such a boost does not occur
in real clusters, where metallicity gradients are much shallower and nowadays
so well resolved, that the overall mass weighted metallicity can be 
reconstructed with confidence --- and is found not
significantly different from the emission weighted one (DELM).
Since the meaningful physical quantity is, in the end, the actual amount 
of metals in the ICM, we prefer to discuss the mass-weighted metallicity, 
in simulations and in observations.}

\begin{figure}
\includegraphics[width=85mm]{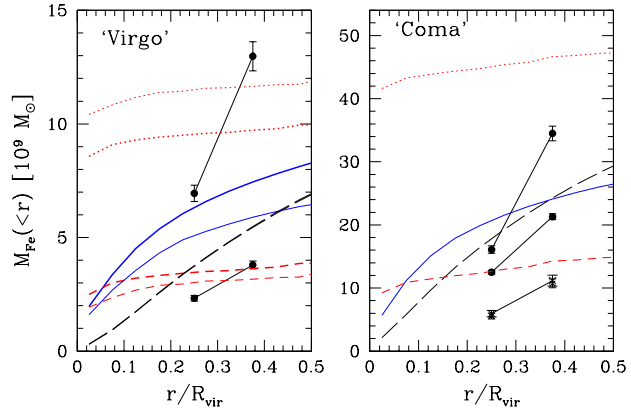}
  \caption{Cumulative Iron mass distributed in stars ({\it red, dotted}:
uncorrected; {\it dashed}: corrected for the cooling flow stars since
$z=2$) 
and in the ICM ({\it blue, solid}), for the ``Virgo'' (thin line at normal
and thick at high resolution) and ``mini-Coma'' AY-SW simulations.
{\it Long-dashed} is $M_{Fe}^{ICM}$ assuming $\frac{1}{3}$ solar abundance in the ICM
(the two for ``Virgo'' at different resolutions overlap into the same one). 
Data points for ICM Fe mass from De Grandi {\it et al.}\ (2004): cold-core
({\it circles}) 
and non cold-core ({\it crosses}) clusters of size comparable
to the simulated ones: A3526 and 2A0335 Virgo-like; A119 (the only
non--cold-core), A426 and A1795 Coma-like. }
\label{MFeVC}
\end{figure}

We consider the AY-SW simulations of the two richest clusters in our set,
``Virgo'' (at normal and high resolution) and ``Mini-Coma''.
In Fig.~\ref{MFeVC} we show as solid lines the cumulative iron mass 
in the ICM within radius $R$, out to $R_{500}$ (which is $\simeq 0.5 R_{vir}$).
The dashed lines are the expected cumulative profiles assuming for our ICM
gas the typical average iron abundance of 0.3 solar (e.g.\ DELM).
Also plotted are the cumulative iron masses within 
$R_{2500}$ and $R_{1000}$ determined by DELM for individual clusters with 
emission--weighted temperatures close to those of our simulated ``Virgo'' 
and ``Coma'', scaled by a factor $\left( \frac{70}{50} \right)^{-
\frac{5}{2}}$ consistent with the $H_0$ value we adopt. 
There is quite some scatter in the actual iron masses of individual clusters, 
but they nicely bracket the simulation results -- especially
considering the cool core clusters (filled circles).
Out to $R_{500}$, the simulations reproduce the expected values 
(solid vs.\ dashed lines) fairly well.
We note that if the ICM maintains the typical iron abundance of 
$0.3 \, Z_{Fe,\odot}$ beyond
$R_{500}$, then the simulation tend to fall short of the expected
metal content of the ICM; however, the observed metallicity profiles
typically reach less
than $R_{500}$ (e.g.\ Fig.~\ref{FehVC}) so that some extrapolation is
already involved in Fig.~\ref{MFeVC}.

Observationally the ``partition'' of the iron mass is skewed in favour 
of the ICM (Renzini et al. 1993; Renzini 1997, 2004); for the AY IMF
in particular, the expected partition is about twice more iron in the ICM 
than in stars (Portinari 2005).
In Fig.~\ref{MFeVC} instead, the partition is skewed in favour of the stars 
(solid vs.\ dotted lines), so that only about 1/3 of the total iron content 
remains in the ICM. The discrepancy may be partly due to the fact that
we compute the iron mass from all the star particles (including the cD, 
its envelope and the intracluster stars), while observational estimates are
based only on galaxies. In very relaxed clusters similar to our simulated ones,
the amount of iron contained in the envelope of the cD can be significant; 
if observationally this is not accounted for, one underestimates the actual 
star--to--ICM iron mass ratio.

However we find that most of the discrepancy is just an artifact due
to significant production and lock-up of iron
by stars formed at the base of late
time cooling flows: subtracting the iron content from these stars from
the cumulative stellar iron mass results in the short-dashed line of Fig.~15. 
The mass of the stars involved in the correction is only about 20\%
(e.g.\ Fig.~12, solid vs. open star) but the effects on the locked up
iron mass are major: after the correction, the iron mass partition
becomes about 2/3 in the ICM to 1/3 in the stars, in much better agreement
with the expectations.
\footnote{In principle, also the ICM iron mass should be corrected for the
effects of late time cooling flow induced star formation and iron
recycling, but we find that this effect is less important
and that most of the iron
produced and recycled by cooling flow stars is locked up within this same
spurious population and does not contribute much to enriching the ICM.}
As the prominent cooling flow stellar
population is a non-physical effect (Section 2.2),
the corrected iron masses are to be regarded as much more realistic than
the uncorrected ones. This highlights that a proper modelling of the metal
partition and of the ICM enrichment is strictly related on improved
understanding of the physics of the central regions.

\begin{figure}
\includegraphics[width=70mm]{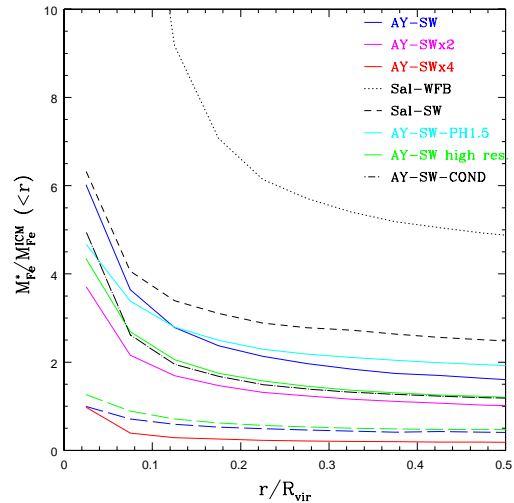}
\caption{Iron star--to--ICM mass ratios for the different simulations
of the ``Virgo'' cluster. As in Fig. 14, the normal and high resolution AY-SW are shown as well, both 
uncorrected ({\it solid}) and corrected since $z$=2 ({\it long-dashed}) for the cooling flow 
stars.}
\label{ratioFe}
\end{figure}

In Fig.~\ref{ratioFe} we plot the star--to--ICM iron mass ratio for different 
runs of the ``Virgo'' cluster, with different physical prescriptions,
to analyse the effects of the different IMFs and feedback schemes on the 
iron partition. Within $R_{500}$ there is typically 2--3 times more iron locked
in the stars than in the ICM, i.e.\ the partition is ``inverted'' with respect
to the expected one as we have seen above.
However, besides the cooling flow corrected Virgo simulations discussed above
(dashed lines), also the AY-SWx2 or AY-SWx4 simulations (with no correction) 
result in reasonable star--to--ICM iron mass ratios: AY-SWx2 yields iron 
equipartition, and AY-SWx4 predicts substantially more iron in the ICM 
than locked in the stellar component. 
One could than wonder whether a mere general enhancement of the feedback 
efficiency could solve the problem of the ``inverted'' partition:
unfortunately the enhanced feedback also suppresses the global star 
formation rate and metal production, so that in spite of the improved 
partition, these simulations yield a lower ICM metallicity than the standard 
AY-SW case and do not match the observed metallicities (Fig.~\ref{FehVC}).

As to the Salpeter simulations, the resulting iron content in the ICM is 
about an order of magnitude below observations, because both the metal production 
and the feedback (i.e.\ metal dispersal) efficiency are too small. 
In the Sal-WFB case the iron distribution is very strongly
skewed towards the stars (ratio 5:1); in the Sal-SW case instead the star--to--ICM iron mass ratio is closer 
to that of the AY-SW case, but due to the lower iron yield of the Salpeter
IMF, both the ICM metallicity (Fig.~\ref{FehVC}) and the stellar metallicity 
(as traced by the colour--magnitude relation of the corresponding cluster 
galaxies, see Fig.~10 in Paper~II) are too low.

\begin{figure}
\includegraphics[width=70mm]{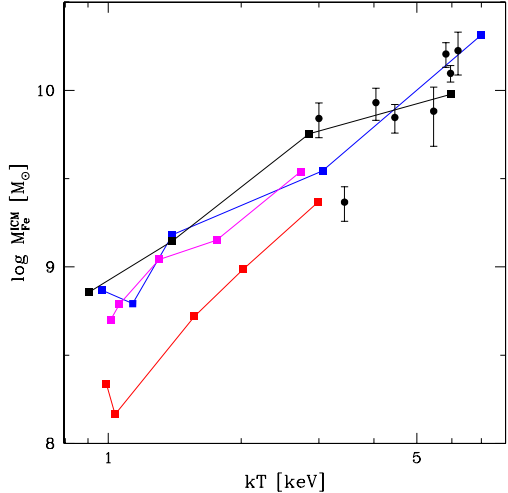}
  \caption{Cumulative Iron masses in the ICM at $R_{2500}$;
models shown as {\it squares}: AY-SW ({\it blue}), AY-SWx2 ({\it magenta}), AY-SWx4 ({\it red}), 
COND. ({\it black}).
{\it Circles}: cold--core clusters from De Grandi {\it et al.} (2004).}
\label{MFeT}
\end{figure}

DELM also point out a correlation between the iron mass within $R_{2500}$ 
(a region well covered by their profiles) and the X--ray luminosity
or temperature. Though few of our simulated clusters overlap with the 
temperature range of the DELM ones, in Fig.~\ref{MFeT} the trend with 
temperature seems in qualitative agreement with the observed one.

\subsection{The Iron Mass-to--Light ratio}

A common way of measuring the efficiency of metal production by cluster
galaxies is the ICM Iron Mass-to-Light Ratio 

\begin{equation}
\label{eq:IMLR}
IMLR_{ICM} \equiv \frac{M^{ICM}_{Fe}}{L_B} = \, < Z^{ICM}_{Fe}> 
\frac{M_{ICM}}{L_B}
\end{equation}
(Ciotti {\it et~al.}\ 1991; Renzini {\it et~al.}\ 1993).
Observational estimates yield 
IMLR$\simeq$0.01-0.02~$M_{\odot}/L_{\odot}$, fairly invariant for
rich clusters above 2--3~keV (Finoguenov {\it et~al.}\ 2000, 2003; DELM).
Such a large IMLR indicates a metal production efficiency higher than
in the local environment by a factor of 3 or more (Portinari {\it et~al.}\ 
2004). The above results based on galactic luminosities might however 
overestimate the actual overall IMLR, and the metal production
efficiency, if the contribution of the intra-cluster stars is significant 
(Lin \& Mohr 2004; Zaritsky, Gonzales \& Zabludoff 2004).

\begin{figure}
\includegraphics[width=70mm]{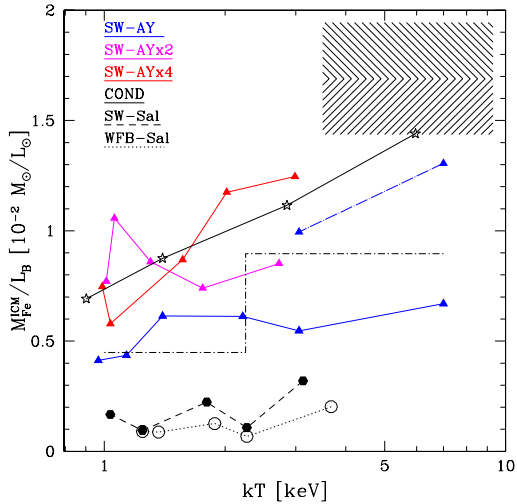}
  \caption{Iron Mass-to-light ratio of the ICM in the {\it B} band, from all 
stars within $R_{500}$. 
{\it Shaded} region: averaged data (along with 
$1\sigma$-scatter) over nine {\it BeppoSAX} nearby rich clusters extrapolated 
to $R_{500}$, from De Grandi {\it et al.}\ (2004). {\it Step curve}: 
average over {\it ASCA} data from Finoguenov {\it et al.} (2003).
The effect of removing stars formed at the base of the cooling flow since $z=2$
is shown as a {\it dot-longdashed} line for the ``Virgo'' and ``Coma'' AY-SW simulations.}
\label{IMLR}
\end{figure}

Fig.~\ref{IMLR} illustrates the IMLR of the simulated clusters 
within $R_{500}$, compared to observational estimates (all scaled
to the adopted value of H$_0$); notice that the {\it BeppoSAX} based 
IMLR of DELM (shaded area) is about twice as large as the {\it ASCA} based 
estimates of Finoguenov {\it et~al.}\ (2003; dash-dotted line).
Though the {\it BeppoSAX} iron profiles are much more accurate, they do not
reach as far out as {\it ASCA} data and significant extrapolation is required 
to derive the cumulative iron mass at $R_{500}$, so that the {\it BeppoSAX} 
values should be regarded as upper limits (DELM).
As to the IMLR of our simulated clusters, the luminosity is computed 
considering all stars (both galactic and inter-galactic) within $R_{500}$:
in our simulations the distribution of stars is highly skewed toward the
cD+intra-cluster stars component, at the expense of the brightest galaxies 
in the luminosity function (Paper~II and~III); hence considering only galactic 
stars in the {\it simulated} clusters would seriously bias the luminosity 
downward. We do however neglect the contribution of the
recent star formation at the centre of our cDs where cooling flows onset
(Section 2.2): 
the effect of dropping the luminosity contribution from these innermost 
young stars is negligible for the lower temperature 
objects, but turns out to be important for the Virgo and Coma-sized ones 
(dot-dashed line in Fig.~\ref{IMLR}). 

The IMLR obtained in simulations can be compatible with the observed one if:
{\it a)} the adopted IMF produces enough metals in the overall; {\it b)} the simulations 
yield the correct partition of metals between the stars and the ICM; 
{\it c)} the star formation history in the simulated cluster is similar 
to that in real clusters, so that the colours and MLR of the simulated 
stellar populations are realistic. 

The Salpeter simulations evidently have a far too low IMLR in the ICM. Since
in principle the Salpeter IMF can produce enough iron to account for the
observed enrichment (though not enough $\alpha$--elements, Portinari
{\it et~al.}\ 2004), the main culprit here is the partition. In the
simulations, there is 3-5 times more iron in the stars than in the ICM
(Fig.~\ref{ratioFe}); if the expected equipartition (Matteucci \& Vettolani
1988; Renzini {\it et~al.}\ 1993; Portinari {\it et~al.}\ 2004) were
reproduced, the iron mass and IMLR of the ICM would be about 3 times larger,
hence compatible with at least the {\it ASCA} results (step line in 
Fig.~\ref{IMLR}). 
The situation as expected improves with the AY simulations, since this 
top--heavy IMF both produces more metals and results in stronger feedback
and metal dispersion into the ICM (even within the same SW feedback 
prescription as the Salpeter case). 

However, simulations are hardly able to reach the very high levels
of IMLR inferred by the {\it BeppoSAX} results. Increasing the feedback
efficiency to SWx2 and SWx4 evidently increases the IMLR in the ICM, because
of the more favourable partition (Fig.~\ref{ratioFe}); nonetheless, as
we remarked in the previous section, even if the right level of IMLR were
reached, the correct ICM metallicity would not necessarily be achieved
in these simulations.

The AY-SW simulations (cooling-flow corrected, with conduction or not)
are confirmed to be those that best reproduce the chemical properties of the
ICM, including the qualitative trend of decreasing IMLR in poor clusters and 
groups ($kT\lsim 2$ keV), as a consequence either of strong galactic winds,
or of lower enriching efficiency.

\subsection{The $\alpha$--elements in the ICM}

Mainly produced by SN~II, $\alpha$--elements are more direct 
tracers of star formation and of the massive stars that have enriched the ICM 
at early times, and also better tracers of the global metallicity ---
which they dominate by mass.
In the ICM $\alpha$--elements are in practice best represented by silicon, 
the second-best measured element in the ICM after iron;
albeit a non--negligible contribution from SNIa, about 80\% of the 
Si production is due to SNII (Portinari et~al.\ 2004). 

The [$\alpha$/Fe] abundance ratios in the ICM of rich clusters are supersolar
and increase outwards, as $\alpha$ elements have a shallower distribution 
than iron (Finoguenov {\it et~al.}\ 2000; B\"ohringer {\it et~al.}\ 2004;
Baumgartner {\it et~al.} 2005).
In particular, they do not mirror the central iron peak in cool core clusters,
which is then imputed to SNIa from the central Brightest Cluster 
Galaxy --- in terms of mass, however, this peak represents only a 10\% 
of the global ICM iron content; De Grandi \& Molendi 2001; DELM; 
B\"ohringer {\it et~al.}\ 2004).

\begin{figure}
\includegraphics[width=70mm]{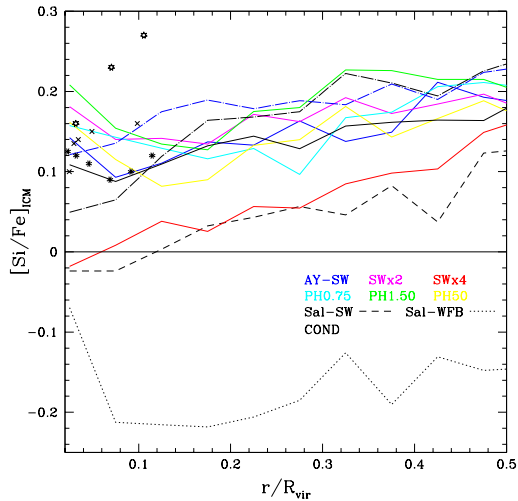}
  \caption{Silicon-over-Iron gradients in the ICM of the two clusters at $z$=0
(``Coma'': {\it dot-dashed}). Line colours as in Fig. 14.
Data from Finoguenov {\it et al.}, 2000: two 
Virgo-sized  clusters (Virgo itself, {\it crosses}; and A1060, {\it asterisks}) 
and the larger one A3112 ({\it stars}).}
\label{SiFehVC}
\end{figure}

Fig.~\ref{SiFehVC} shows that, with the exception of the Sal-WFB simulation
where [Si/Fe] is significantly undersolar everywhere, the trend of supersolar
[Si/Fe] increasing with radius is present in all our simulations. (The upturn
of [Si/Fe] at the very centre in some simulations is due to the recent star
formation in the centre of the cD, and is not significant). All the
simulations AY-SW and AY-SWx2 also reproduce the correct level of [Si/Fe];
while the Sal-SW and extreme feedback AY-SWx4 simulations have much lower 
[Si/Fe]. Notice that, although our feedback and enrichment scheme favours
the ejection and dispersal of SNII products, SNIa still contribute 
significantly to the ICM enrichment: for a Salpeter or AY IMF, 
pure SNII enrichment would correspond to [Si/Fe]$\sim$+0.5, 
hence, though supersolar, the level of [Si/Fe] in the simulations indicates
a significant contribution from SNIa. Actually, for the
AY IMF the expected [Si/Fe] for the overall enrichment, including all SNII and
SNIa, is [Si/Fe]=+0.2~dex (supersolar since the AY IMF is top--heavy,
favouring SNII in proportion to SN~Ia); henceforth, for the AY-SW 
simulations the ICM is enriched by SNII and SNIa in roughly the same
proportion as they are in the corresponding IMF.

We can compare our results with those of Tornatore {\it et~al.}\
(2004): their simulated cluster is comparable in virial mass to our 
``Virgo'', and they also present Salpeter and AY IMF simulations 
with different feedback efficiencies; but their ``sub-grid'' treatment 
of star formation and feedback differs from ours. They share our same problem 
of too steep iron gradients, in all cases; but their abundance ratio profiles 
behave quite differently from those in Fig.~\ref{SiFehVC}.
Their [$\alpha$/Fe] is roughly constant, solar or undersolar, 
over most of the cluster; probably they find lower [$\alpha$/Fe] than
we do because their feedback prescription treats SNII and SNIa
alike, while ours favours the diffusion of SNII products.
Moreover, in the central regions ($R \leq 0.1 R_{200}$) 
they find a peak in [$\alpha$/Fe] due to recent star formation and 
SNII explosions; indeed their Salpeter models A and B with ``standard'' 
wind parameters display profiles very similar to our Sal-WFB simulation. 
In all our SW simulations, instead,
we find at most a mild central enhancement in [$\alpha$/Fe]: our SW feedback 
prescription seems more efficient at contrasting late star formation 
at the centre of the cooling flow.
Tornatore {\it et~al.} are able to eliminate such central [$\alpha$/Fe] spikes 
when invoking additional AGN energy input (Model C), but at the expense 
of retarding significantly the star formation history 
--- which is not the case in our SW simulations (Paper~II).
One is then tempted to conclude that their ``standard'' feedback
implementation is as little efficient as our WFB case, and they need to
resort to ``AGN powering'' to eliminate the central spike while we can do it
with the sole SNII energy in the SW simulation. But things are not so
straightforward: in their model D (AY IMF with ``standard'' wind paramenters)
the high ICM metallicity and [$\alpha$/Fe]$\sim$+0.3
suggest a very efficienct dispersal of metals in the ICM, more than
in our AY-SW simulations; their feedback implementation now looks more
efficient than our SW one, hence it seems that their scheme is 
much more sensitive to the underlying IMF. 
Though more detailed discussion on different implementations is out of the 
scopes of this paper, we stress the importance of having different algorithms 
and codes to assess the subtle but important effects of sub-grid physics.

\subsection{The redshift evolution of the ICM metallicity}

\begin{figure}
\includegraphics[width=80mm]{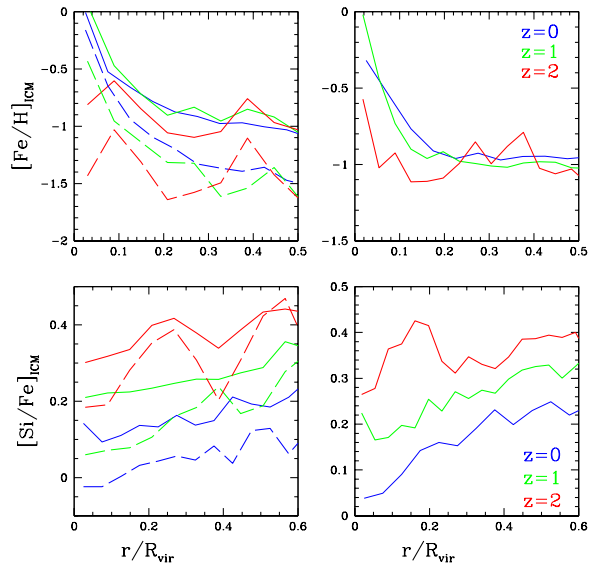}
\caption{Radial profiles of Iron ({\it top}) and Silicon ({\it bottom}) 
abundances in the ICM at z=0,1,2. {\it Left}: ``Virgo'' ({\it solid}: AY-SW, 
{\it dashed}: Sal-SW); {\it right}: ``Coma''. $R_{vir}$ evaluated at z=0.}
\label{SiFez}
\end{figure}

There is no significant variation in the typical ICM iron abundance of
clusters up to $z \simeq 1.2$, implying that the bulk of the enrichment
occurs at earlier epochs (Mushotzky \& Loewenstein 1997; Donahue {\it et~al.}\ 
2003; Tozzi {\it et~al.}\ 2003).
Fig.~\ref{SiFez} (top panels) shows that the overall iron abundance, and 
its profile, in the ICM of simulated clusters is essentially unchanged 
from $z=0$ to $z=1$, in very good agreement with observations.
Any evolution in the ICM metallicity at $z<1$ is limited to the
very central regions, $r<0.1 R_{vir}$, and is due to the late
cooling flows and star formation at the centre of the cD; excluding
this effect, 
since the bulk of the stars in the cluster are formed at $z \gsim$2 (Papers~II
and~III), at $z$=2 we predict somewhat lower iron metallicities,
but such redshift range is not probed by observations yet.
The lower panels in the figure show the evolution in [Si/Fe], which as expected
decreases in time due to the delayed iron contribution of SNIa 
with respect to the bulk of the silicon production (from SNII).
The change in abundance ratios from $z=0$ to $z=1$ is minor, about
0.1~dex.

There are indications that some evolution in the metallicity of the ICM
could be expected due to ram-pressure stripping of the metal richer cold
gas from galaxies (Domainko \etal 2006). However, further study of the redshift 
evolution of metallic abundances and gradients
in the ICM (and in the intra-cluster light as well: see Paper III) out
of our simulated data is presently on progress.

\section{Conclusions}

All the results reported are based on cosmological
simulations of galaxy clusters including self-consistently
cooling, star formation, supernova driven galactic 
super-winds, chemical evolution, UV field and thermal conduction, as
described in Section 2.
In relation to modelling the properties of cluster gas and galaxies,
especially from the chemical point of view, this 
represents an important step forward with respect to 
previous theoretical works on the subject (see also Paper II and III).

In Section 3 the gas distribution has been analysed, in terms of cumulative
fractions of baryonic, gas and stellar components. When compared with the
``adiabatic" model (no cooling/star formation), the combined effect of
cooling and feedback in our simulations makes the gas fraction shift 
significantly down at all distances. Enhancing feedback, the fraction
of gas turned into stars decreases, thus augmenting the cumulative gas
fraction up to the virial radius; and at the same time more gas is spread
out at larger distances, resulting into a more extended profile (Fig. 1)
-- this effect becoming even more evident when looking at smaller systems
(Fig. 3). From comparing gas distributions reached at $R_{500}$ in systems
of different size, we find that feedback and cooling combined together 
again introduce a trend, with respect to the ``adiabatic" run, which is increasing
with temperature and which is more prominent in the stronger feedback and
in the pre-heating models (Fig. 4); this in turn drives the baryon fraction--
temperature correlation, as the stellar fraction seems to follow a fairly
constant behaviour which might be compatible with a similarity of the star
formation process in systems over a significant mass range.

In modelling the temperature profiles (Section 4), fair agreement has been found
with observed distribution from $\sim 0.25R_{vir}$ outwards,
while an extended isothermal core 
and the positive gradient
in the innermost regions ($r<0.1 R_{vir}$) are still lacking in
most models; feedback and pre--heating help to smooth out the steep
negative temperature gradient typical of simulations with cooling,
yet the temperature profile remains more sharply peaked than 
the observed one (Fig. 5).
Thermal conduction is found to affect the higher temperature,
more massive systems in such a way that outer radial temperature gradients are
reduced and ``cold fronts" are smoothed out.

As for the entropy profiles, we needed to appeal to the most extreme
SWx4 scheme in order to lower the central density to values consistent
with the presence of an entropy-floor at 
$\sim 100$keV$\cdot$cm$^2$, and even higher in the smaller-scales
simulated systems.
The observed level and slope of the $L_X-T$ relation are also best 
reproduced by strenghtening the feedback up to the SWx4 case, at least
in the probed range 1-6 keV; otherwise, the low-entropy gas at the
cluster's centre induces over-emissivity, as seen in other simulations. 
Although general results from temperature and entropy distributions 
still call for the need of
balancing cooling by enhancing feedback, it is however 
a firm result that our AY-SW models can actually fit cold-core
systems, reproducing observed values of central entropy at 10-20 
keV$\cdot$cm$^2$.

The findings on the mass-to-light ratio of the ICM and the cold fraction 
place both in a fair agreement with observations, when adopting the
standard SW scheme for stellar feedback
with a top-heavy IMF, while a Salpeter IMF yields too large
stellar fractions and masses, giving rise to an excess optical luminosity
--as it is evident from the radial profiles (Fig. 13).
The cold fraction scales with cluster size, decreasing as system mass
or temperature increases, as observed; 
in the simulations this trend is due to the
more extended gas distribution, and the smaller gas and baryonic 
fraction within $R_{500}$, characterizing systems of lower mass
and temperature, rather than to an increase in the efficiency 
of galaxy formation and in the stellar fraction.

In Section 8 the metal enrichment of the ICM has been analysed. It
has been found that the iron abundance profiles show a steep inner gradient,
and an iron abundance overall level which is somewhat low relative
to observations, at least at $r\ga 0.1 R_{vir}$.
The steep inner gradient, along with the high iron level within this inner part,
is mostly due to the excessive star formation, with subsequent chemical evolution, 
resulted from the spurious, late-time central cooling flows. This
problem would be cured, if e.g. some non-supernova related
feedback mechanisms could be invoked for halting the cooling flows
(see below).

Regarding the overall level of iron abundance, it is clear from our results that
adopting a top-heavier IMF such as Arimoto-Yoshii rather than a Salpeter one,
helps to improve the distribution considerably. 
Enhancing the feedback then helps to balance the
partition between metals in stars and in the ICM, 
favouring a higher iron 
MLR of the ICM, but this comes at the expense of hampering
star formation and metal production itself, resulting in
a too low metallicity for the ICM. Rather than a stronger
stellar feedback, an improved understanding of the cooling 
and heating processes in the very central regions and of the
formation of the cD seems to be the key to better 
balancing the total iron content of the stars and ICM.
In fact, when excluding from the stellar iron mass
the iron produced by stars 
formed at the base of the cooling flow, 
the metal partition between stars and ICM
gets skewed in favour of the latter, thus much closer to empirical estimates.
Then realistic values of metal enrichment through the ICM are attained,
as seen from the cumulative iron masses in ``Coma'' and ``Virgo''
modelled and well reproduced by AY-SW scheme (Fig. 15). And at the same time,
such cooling flow correction helps the cold fraction as well to get reduced 
to values lower than 20\%, excellently matching those observed (Figs. 12-13).

Besides iron, the distribution of silicon as tracer of $\alpha$-elements
from SNII in the ICM shows a non-constant, increasing radial trend with respect
to iron abundance, where 
the AY-SW model is fairly efficient in preventing the central peak due to
recent star formation, as is clear from comparison with the WFB case (Fig. 19).
Although the evolution of the thermal and chemical properties of the ICM over
significant redshift ranges will be in perspective treated in another forthcoming paper,
here we just stress the attention upon the behaviour of gas metal
abundances since $z$=2: iron levels appear to remain essentially unchanged at
least from $z$=1 to the present; silicon abundance with respect to iron
instead decreases due to the delayed contribution of SNIa (Fig. 20).

In summary, although we can reproduce many observed features of groups and 
clusters, our results point out the need of a
general heating mechanisms which could be able to suppress the
overproduction of stars at late times in the central cluster regions,
without affecting the overall cluster star formation. 
Hot unstable buoyant bubbles as observed in X--ray (e.g. Ensslin \&
Heinz, 2002
for a model), released in association with AGN jets, can transport
energy by convection, thus providing a means of keeping the central
gas from cooling. Another possibility is that the ICM might contain
random magnetic fields of sufficient strength to significantly reduce
the gas inflow by the combined effect of magnetic pressure and 
reconnection. 
Yet another option could be
dissipation of sound waves resulting from AGN induced, turbulent motions,
as recently proposed by Fujita \& Suzuki (2005).


\section*{Acknowledgments}
We thank S. De Grandi, J. Rasmussen and A. Sanderson for having provided 
us with their observational data; S. Borgani for fruitful discussions;
and the referee E. van Kampen for precious comments.

This work was funded by the EU through the European Structural Social Fund 
/ P.O.N. 2000-2006,
by Danmarks Grundforskningsfond through its support
for the establishment of the (now expired) TAC--Theoretical Astrophysics Center,
by NORDITA, by the Villum Kann Rasmussen Foundation 
and by the Academy of Finland (grant nr.~208792). 
The Dark Cosmology Centre is funded by the DNRF.
All computations reported in this paper were performed on the IBM SP4
and SGI Itanium II facilities provided by Danish Centre for Scientific
Computing (DCSC).


\begin{thebibliography}{99}
\bibitem[2003]{A.03} Abadi M.G., Navarro J.F., Steinmetz M., Eke V.R., 2003, ApJ 591, 499
\bibitem[2001]{Aguirre} Aguirre A., Hernquist L., Schaye J., Katz N., 
Weinberg D.H., Gardner J., 2001, ApJ 561, 521
\bibitem[1998]{} Allen S.W., Fabian A.C., 1998, MNRAS 297, L57
\bibitem[2003]{Allen} Allen S.W., Fabian A.C., Johnstone R.M., Arnaud K.A.,
Nulsen P.E.J., 2001, MNRAS 322, 589

\bibitem[\protect\citeauthoryear{Antonuccio et al.}{2003}]{} 
Antonuccio-Delogu V., Becciani U., Ferro D., 2003, Comput. Phys. Commun. 155, 159
\bibitem[]{} Arimoto N., Yoshii Y., 1987, A\&A 173, 23

\bibitem[\protect\citeauthoryear{Arnaud {\it et~al.}}{1992}]{1992A&A...254..49}
Arnaud M., Rothenflug R., Boulade O., Vigroux L., Vangioni--Flam E., 
1992, A\&A 254, 49
\bibitem[\protect\citeauthoryear{Arnaud and Evrard}{1999}]{1999MNRAS.305..631A}
Arnaud M., Evrard A.E., 1999, \mnras 305, 631 	
\bibitem[\protect\citeauthoryear{Arnaud {\it et~al.}}{2001}]{2001A&A.365..L67} 
Arnaud M., et~al.\ 2001, A\&A 365, L67
\bibitem[\protect\citeauthoryear{Babul et al.}{2002}]{} Babul A., Balogh M.L.,
Lewis G.F., Poole G.B., 2002, MNRAS 330, 329
\bibitem[\protect\citeauthoryear{Balogh et al.}{1999}]{1999MNRAS.307..463B} 
Balogh M.L., Babul A., Patton D.R., 1999, \mnras 307, 463	
\bibitem[\protect\citeauthoryear{Balogh et al.}{2001}]{2001MNRAS.326.1228B} 
Balogh M.L., Pearce F.R., Bower R.G., Kay S.T., 2001, \mnras 326, 1228	
\bibitem[2004]{Baum05} Baumgartner W.H., Loewenstein M., Horner D.J., 
Mushotzky R.F., 2005, ApJ 620, 680
\bibitem[2001]{BdJ} Bell E.F., de Jong R.S., 2001, \apj 550, 212
\bibitem[]{} Bialek J.J., Evrard A.E., Mohr J.J., 2001, ApJ 555, 597
\bibitem[2004]{Bohr04} B\"ohringer H., Matsushita K., Churazov E., 
Finoguenov A., Ikebe Y., 2004, A\&A 416, L21

\bibitem[\protect\citeauthoryear{Borgani et al.}{2001}]{2001ApJ...559L..71B} 
Borgani S., et al.,
2001, \apj 559, 71
\bibitem[\protect\citeauthoryear{Borgani et al.}{2002}]{2002MNRAS.336..409B} 
Borgani S., et al.,
2002, \mnras 336, 409
\bibitem[\protect\citeauthoryear{Borgani et al.}{2004}]{2004MNRAS.348.1078B}
Borgani S., et al., 2004, MNRAS 348, 1078

\bibitem[1997]{Bower1} Bower R.G., 1997, MNRAS 288, 355
\bibitem[1997]{Bower2} Bower R.G., Benson A.J., Lacey C.G., Baugh C.M., 
Cole S., Frenk C.S., 2001, MNRAS 325, 497
\bibitem[1998]{BT98} Boyle B.J., Terlevich R.J., 1998, MNRAS 239, L49
\bibitem[2001]{} Brighenti F., Mathews W.G., 2001, ApJ 553, 103 

\bibitem[]{} Carilli C.L., Taylor G.B., 2002, ARofA\&A 40, 319
\bibitem[]{beta} Cavaliere A., Fusco-Femiano R., 1976, A\&A 49, 137
\bibitem[]{Cavaliere1} Cavaliere A., Menci N., Tozzi P., 1997, ApJ 484, L21
\bibitem[]{Cavaliere2} Cavaliere A., Menci N., Tozzi P., 1998, ApJ 501, 493
\bibitem[]{Cavaliere4} Cavaliere A., Lapi A., Menci N., 2002, ApJ 581, L1
\bibitem[]{} Cen R., Ostriker J.P., 1994, ApJ 429, 4

\bibitem[1997]{CO97} Ciotti L., Ostriker J.P., 1997, ApJ 487, L105
\bibitem[2001]{CO2001} Ciotti L., Ostriker J.P., 2001, ApJ 551, 131
\bibitem[1991]{Cio91} Ciotti L., D'Ercole A., Pellegrini S., Renzini A., 1991, ApJ 376, 380

\bibitem[]{} Cleary P.W., Monaghan J.J., 1990, ApJ 349, 150
\bibitem[]{} Cowie L.L., McKee C.F., 1977, ApJ 211, 135
\bibitem[]{} Croton D.J., Springel V., White S.D.M., et al., 2006, MNRAS 365, 11

\bibitem[]{} Dav\`e R., et al.,
2001, ApJ 552, 473
\bibitem[]{} Dav\`e R., Katz N., Weinberg D.H., 2002, ApJ 579, 23

\bibitem[1992]{David90} David L.P., Arnaud K.A., Forman W., Jones C., 
1990, ApJ 356, 32
\bibitem[]{David92} David L.P., Jones C., Forman W., 1992, ApJ 388, 82
\bibitem[]{David93} David L.P., Slyz A., Jones C., Forman W., Vrtilek S.D., 
Arnaud K.A., 1993, ApJ 412, 479
\bibitem[]{David95} David L.P., Jones C., Forman W., 1995, ApJ 445, 578

\bibitem[\protect\citeauthoryear{De Grandi et al.}{2003}]{} 
De Grandi S., Ettori S., Longhetti M., Molendi S., 2004, A\&A 419, 7 (DELM)
\bibitem[\protect\citeauthoryear{De Grandi and Molendi}{2001}]{2001ApJ...551..153D} 
De Grandi S. and Molendi S., 2001, \apj 551, 153					
\bibitem[\protect\citeauthoryear{De Grandi and Molendi}{2002}]{} De Grandi S., 
Molendi S., 2002, \apj 567, 163		

\bibitem[1986]{DS86} Dekel A., Silk J. 1986, ApJ 303, 39
\bibitem[]{} Domainko W., Gitti M., Schindler S., Kapferer W., 2004, A\&A 425,
L21 						
\bibitem[]{} Domainko W., et al.,
2005, A\&A 452, 795

\bibitem[]{} Donahue M. and ROXS Collaboration, 1999, ApJ 527, 525
\bibitem[2003]{Donahue03} Donahue M., Gaskin J.A., Patel S.K., Joy M., 
Clowe D., Hughes J.P., 2003, ApJ 598, 190
\bibitem[2005]{DOnghia}
D'Onghia E., Sommer--Larsen J., Romeo A.D., Burkert A., Pedersen K., 
Portinari L., Rasmussen J., 2005, ApJL 630, L109
\bibitem[]{} Edge A.C, Stewart G.C., 1991, MNRAS 252, 414
\bibitem[Eisenstein \& Hut(1998)]{1998ApJ...498..137E} Eisenstein D.~J.,
\& Hut P., 1998, \apj 498, 137
\bibitem[]{} Eke V.R., Navarro J.F., Frenk C., 1998, Apj 503, 569
\bibitem[\protect\citeauthoryear{Eldridge and Tout}{2004}]{ET04} 
Eldridge J.J., Tout C.A., 2004, MNRAS 353, 87
\bibitem[]{} Ensslin T.A., Heinz S., 2002, A\&A 384, 27

\bibitem[\protect\citeauthoryear{Ettori et al.}{2002}]{} Ettori S., De Grandi S, 
Molendi S., 2002, A\&A 391, 841		
\bibitem[]{} Ettori S., Fabian A.C., 2000, MNRAS 317, 57
\bibitem[\protect\citeauthoryear{Evrard1}{1990}]{1990ApJ...363..349E} Evrard A.E., 
1990, ApJ 363, 349
\bibitem[\protect\citeauthoryear{Evrard2}{1996}]{} Evrard A.E., Metzler Ch.A.,
Navarro J.F., 1996, ApJ 469, 494
\bibitem[\protect\citeauthoryear{Evrard3}{1991}]{} Evrard A.E., Henry J.P., 1991, 
ApJ 383, 95
\bibitem[1994]{CF} Fabian A.C., 1994, ARofA\&A 32, 277

\bibitem[]{} Finoguenov A., Ponman T.J., 1999, MNRAS 305, 325		
\bibitem[\protect\citeauthoryear{Finoguenov et al.}{2000}]{} Finoguenov A., David L.P., 
Ponman T.J., 2000, ApJ 544, 188 		
\bibitem[\protect\citeauthoryear{}{}]{} Finoguenov A., Jones C., Böhringer H., Ponman T.J., 
2002, ApJ 578, 74							
\bibitem[\protect\citeauthoryear{Finoguenov et al.}{2003}]{2003ApJ...594..136F} 
Finoguenov A., Burkert A., B\"ohringer H., 2003, \apj 594, 136		

\bibitem[]{} Franceschini A. et al., 2003, MNRAS 343, 1181
\bibitem[]{} Frenk C.S. et al., 1999, ApJ 525, 554

\bibitem[]{} Fujita Y., Suzuki T.K., 2005, ApJ 630, L1
\bibitem[1998]{Fuka98} Fukazawa Y., Makishima K., Tamura T., Ezawa H., 
Xu H., Ikebe Y., Kikuchi K., Ohashi T., 1998, PASJ 50, 187

\bibitem[1997]{GSL} Gelato S., Sommer-Larsen J., 1999, MNRAS 303, 321
\bibitem[2001]{Ger01} 
Gerhard O., Kronawitter A., Saglia R.P., Bender R., 2001, AJ 121, 1936
\bibitem[1997]{G97} Gerritsen J.P.E., 1997, Ph.D. thesis, Kapetyn Astron. Inst.
\bibitem[\protect\citeauthoryear{Governato et al.}{2004}]{G.04} Governato F. et al. 2004, ApJ 607, 688

\bibitem[1996]{UV} Haardt F., Madau P., 1996, ApJ 461, 20
\bibitem[\protect\citeauthoryear{Helsdon and Ponman}{2000a}]{2000MNRAS.315..356H} 
Helsdon S.F., Ponman T.J., 2000a, \mnras 315, 356		
\bibitem[\protect\citeauthoryear{Helsdon and Ponman}{2000b}]{2000MNRAS.319..933H} 
Helsdon S.F., Ponman T.J., 2000b, \mnras 319, 933		

\bibitem[]{} Jubelgas M., Springel V., Dolag K., 2004, MNRAS 351, 423
\bibitem[1991]{Kaiser} Kaiser N., 1991, ApJ 383, 104

\bibitem[]{} Kapferer W., et al., 2006, A\&A 447, 827
\bibitem[]{} Katz N., White S.D.M., 1993, ApJ 412, 455
\bibitem[]{} Kay S.T., Thomas P.A., Theuns T., 2003, MNRAS 343, 608
\bibitem[]{} Kay S.T., Thomas P.A., Jenkins A., Pearce F.R., 2004, MNRAS 355, 1091
\bibitem[]{} Knight P.A., Ponman T.J., 1997, MNRAS 289, 955
\bibitem[]{} Kravtsov A.V., Nagai D., Vikhlinin A.A., 2005, ApJ 325, 588
\bibitem[]{} Kravtsov A.V., Yepes G., 2000, MNRAS 318, 227
\bibitem[]{} Larson R.B., 1974, MNRAS 169, 229
\bibitem[]{} Lewis G.F., Babul A., Katz N., Quinn Th., Hernquist L., 
Weinberg D.H., 2000, ApJ 536, 623

\bibitem[\protect\citeauthoryear{Lia et al.}{2002a}]{2002MNRAS.330..821L} 
Lia C., Portinari L., Carraro G., 2002, \mnras 330, 821 

\bibitem[\protect\citeauthoryear{Lin et al.}{2003}]{2003ApJ...591..749L} 
Lin Y., Mohr J., Stanford S.A., 2003, \apj 591, 749 (LMS)		
\bibitem[\protect\citeauthoryear{Lin \& Mohr}{2004}]{LM2004} 
Lin Y.-T., Mohr J.J., 2004, ApJ 617, 879

\bibitem[]{} Lloyd-Davies E.J., Ponman T.J., Cannon D.B., 2000, MNRAS 315, 689
\bibitem[]{} Loewenstein M., Mushotzky R.F., 1996, ApJ 466, 695
\bibitem[2001]{Loe01} Loewenstein M., 2001, ApJ 557, 573

\bibitem[]{} Lou Y., Bian F., 2005, MNRAS 358, 1231
\bibitem[1999]{MF99} Mac Low, M.-M., Ferrara, A. 1999, ApJ 513, 142
\bibitem[\protect\citeauthoryear{Markevitch}{1998}]{1998ApJ...504...27M} 
Markevitch M., 1998, \apj 504, 27		
\bibitem[]{} Markevitch M. et al., 2000, ApJ 541, 542
\bibitem[]{} Matteucci F., Vettolani G., 1988, A\&A 202, 21
\bibitem[]{} McCarthy I.G., Balogh M.L., Babul A., Poole G.B., Horner D.J.,
2004, ApJ 613, 811
\bibitem[]{} McNamara B.R. et al., 2000, ApJ 534, L135
\bibitem[]{} Menci N., Cavaliere A., 2000, MNRAS 311, 50
\bibitem[1994]{ME94} Metzler C.A., Evrard A.E., 1994, ApJ 437, 564
\bibitem[]{MM} Mo H. J., Mao S., 2002, MNRAS 333, 768
\bibitem[1985]{ML85} Monaghan J. J., Lattanzio J. C. 1985, A\&A 149, 135
\bibitem[]{} Moretti A., Portinari L., Chiosi C., 2003, \aap 408, 431

\bibitem[1997]{M.97} Mori M., Yoshii Y., Tsujimoto T., Nomoto, K.  1997, ApJ 478, L21
\bibitem[1984]{Mushotzky} Mushotzky R.F., 1984, PhScrT 7, 157
\bibitem[1997]{ML97} Mushotzky R.F., Loewenstein M., 1997, ApJ 481, L63
\bibitem[]{} Nagai D., Kravtsov A.V., 2004, in Outskirts 
of galaxy clusters, IAU Coll.\ 195, eds.\ 
A.~Diaferio et~al., p.296
\bibitem[]{} Narayan R., Medvedev M.V., 2001, ApJ 562, L129

\bibitem[1995]{NFW} Navarro J.F, Frenk C., White S.D.M., 1995, MNRAS 275, 720
\bibitem[]{} Pearce F.R., Thomas P.A., Couchman H.M.P., Edge A.C., 2000,
MNRAS 317, 1029

\bibitem[]{} Persic M., Rephaeli Y., 2002, A\&A 382, 843
\bibitem[Piffaretti et al.(2005)]{2005AA...433..101P} Piffaretti R.,
 Jetzer P., Kaastra J.~S., Tamura T., 2005, A\&A 433, 101	

\bibitem[]{Pipino} Pipino A., Matteucci F., Borgani S., Biviano A., 2002, 
NewA 7, 227
\bibitem[]{} Ponman T.J., Bourner P.D.J., Ebeling H., Bohringer H., 1996, MNRAS 283, 690
\bibitem[]{} Ponman T.J., Cannon D.B., Navarro J.F., 1999, Nature 397, 135	
\bibitem[\protect\citeauthoryear{Ponman et al.}{2003}]{2003MNRAS.343..331P} 
Ponman T.J., Sanderson A.J.R., Finoguenov A., 2003, \mnras 343, 331		

\bibitem[\protect\citeauthoryear{Portinari et al.}{1998}]{1998} 
Portinari L., Chiosi C., Bressan A., 1998, A\&A 334, 505
\bibitem[\protect\citeauthoryear{Portinari et al.}{2004}]{2004} 
Portinari L., Moretti A., Chiosi C., Sommer--Larsen J., 2004, ApJ 604, 579
\bibitem[2004]{Port04} Portinari L., 2005, in Modelling the Intergalactic and 
Intracluster Media,
V.\ Antonuccio et~al.\ (eds.), MemSAIt Suppl.\ in press (astro-ph/0402486)

\bibitem[]{} Pratt G.W., Arnaud M., 2003, A\&A 408, 1

\bibitem[]{} Ranalli P., Comastri A., Setti G., 2003, A\&A 399, 39
\bibitem[\protect\citeauthoryear{Rasmussen and Ponman}{2004}]{} Rasmussen J., 
Ponman T.J., 2004, MNRAS 349, 722
\bibitem[\protect\citeauthoryear{Rasmussen et al.}{2004}]{} Rasmussen J., 
Sommer-Larsen J., Toft S., Pedersen K., 2004, MNRAS 349, 255

\bibitem[1997]{Recchi97} Recchi S., Matteucci F., D'Ercole A., 2001, 
MNRAS 322, 800
\bibitem[\protect\citeauthoryear{Renzini1}{1997}]{} Renzini A., 1997, ApJ 488, 35
\bibitem[]{} 
Renzini A., 2004, in Clusters of galaxies,
ed.\ J.S.~Mulchaey, A.~Dressler \& A.~Oemler (Cambridge University Press), p.~260
\bibitem[]{} 
Renzini A., Ciotti L., D'Ercole A., Pellegrini S., 1993, ApJ 419, 52
\bibitem[]{} 
Ritossa C., Garcia-Berro E., Iben I.Jr., 1996, ApJ 460, 489
\bibitem[]{} 
Ritossa C., Garcia-Berro E., Iben I.Jr., 1999, ApJ 515, 381
\bibitem[\protect\citeauthoryear{Robertson et al.}{2004}]{R.04}
  Robertson B., Yoshida N., Springel V., Hernquist L. 2004, ApJ 606, 34
\bibitem[2002]{Romano2002} 
Romano D., Silva L., Matteucci F., Danese L., 2002, MNRAS 334, 444

\bibitem[\protect\citeauthoryear{PaperII}{2005}]{PII}
Romeo A.D., Portinari L., Sommer-Larsen J., 2005, MNRAS 361, 983 (Paper II)

\bibitem[\protect\citeauthoryear{Roussel et al.}{2000}]{2000A&A...361..429R} 
Roussel H., Sadat R., Blanchard A.,
2000, \aap 361, 429
\bibitem[1955]{Salpeter} Salpeter E.E., 1955, ApJ 121, 161

\bibitem[\protect\citeauthoryear{Sanderson et al.}{2003}]{2003MNRAS.340..989S} 
Sanderson A.J.R., Ponman T.J.,
Finoguenov A., Lloyd-Davies E.J., Markevitch M., 2003, \mnras 340, 989	

\bibitem[]{} Scannapieco C., Tissera P.B., White S.D.M., Springel V., 2005,
MNRAS 364, 552

\bibitem[]{} Schindler S., et al., 2005, A\&A 435, L25
\bibitem[1998]{SR98} Silk J., Rees M. J. 1998, A\&A 331, L1

\bibitem[\protect\citeauthoryear{Sommer-Larsen et al.}{2003}]{2003ApJ...596...47S} 
Sommer-Larsen J., G\"otz M., Portinari L., 2003, \apj 596, 47 (SLGP)
\bibitem[\protect\citeauthoryear{PaperIII}{2005}]{PIII} Sommer-Larsen J.,
Romeo A.D., Portinari L., 2005, MNRAS 357, 478 (Paper III)

\bibitem[\protect\citeauthoryear{Springel and Hernquist}{2002}]{SH02} Springel V., 
Hernquist L., 2002, \mnras 333, 649
\bibitem[2004]{SDMH04} Springel V., Di Matteo T., Hernquist L., 2005, ApJ 620, L79

\bibitem[]{} Suginohara T., Ostriker J.P., 1998, ApJ 507, 16
\bibitem[]{} Sutherland R.S., Dopita M.A., 1993, ApJS 88, 253 (SD)

\bibitem[]{} 
Tamura T., Kaastra J.S., den Herder J.W.A., Bleeker J.A.M., Peterson J.R.,
2004, A\&A 420, 135
\bibitem[2000]{TC00} Thacker R. J., Couchman H. M. P. 2000, ApJ 545, 728
\bibitem[2001]{TC01} Thacker R. J., Couchman H. M. P. 2001, ApJ 555, L17
\bibitem[\protect\citeauthoryear{Thomas \& Couchman}{1992}]{1992MNRAS.257...11T}
Thomas P.A., Couchman H.M.P., 1992, MNRAS 257, 11
\bibitem[2002]{Thomas et al.} Thomas P.A., Muanwong O., Kay S.T., Liddle A.R.,
2002, MNRAS 330, L48

\bibitem[\protect\citeauthoryear{Tornatore et al.}{2003}]{2003MNRAS.342.1025T} 
Tornatore L., Borgani S., Springel V., Matteucci F., Menci N., Murante G., 2003, \mnras 342, 1025
\bibitem[\protect\citeauthoryear{Tornatore et al.}{2004}]{Torna04} 
Tornatore L., Borgani S., Matteucci F., Recchi S., Tozzi P., 
2004, \mnras 349, L19
\bibitem[]{} Tozzi P., Norman C., 2001, ApJ 546, 63
\bibitem[]{} 
Tozzi P., Rosati P., Ettori S., Borgani S., Mainieri V., Norman C., 2003,
ApJ 593, 705

\bibitem[VS1]{} Valageas P., Silk J., 1999a, A\&A, 347, 1
\bibitem[VS2]{} Valageas P., Silk J., 1999b, A\&A, 350, 725
\bibitem[\protect\citeauthoryear{Valdarnini}{2003}]{2003MNRAS.339.1117V} 
Valdarnini R., 2003, \mnras 339, 1117

\bibitem[]{} Vikhlinin A., Forman W., Jones C., 1999, ApJ 525, 47		

\bibitem[]{} Voigt L.M., Fabian A.C., 2004, MNRAS 347, 1130
\bibitem[]{} Voigt L.M., Schmidt R.W., Fabian A.C., Allen S.W., Johnstone
R.M., 2002, MNRAS 335, L7
\bibitem[\protect\citeauthoryear{Voit \& Bryan}{2001}]{} Voit G.M., Bryan G.L.,
2001, Nature 414, 425
\bibitem[\protect\citeauthoryear{Voit et al.}{2002}]{} Voit G.M., Bryan G.L., Balogh M.L., Bower R.G.,
2002, ApJ 576, 601								    
\bibitem[\protect\citeauthoryear{Voit et al.}{2003}]{2003ApJ...593..272V} 
Voit G.M., Balogh M.L., Bower R.G., Lacey C.G., Bryan G.L., 2003, \apj 593, 272	    

\bibitem[]{} Yamada M., Fujita Y., 2001, ApJ 553, L145
\bibitem[\protect\citeauthoryear{White et al.}{1993}]{1993Natur.366..429W} 
White S.D.M., Navarro J.F., Evrard A.E., Gnedin N., 1993, Nature 366, 429	

\bibitem[\protect\citeauthoryear{Wu et al.}{2000}]{} Wu K.K.S., Fabian A.C.,
Nulsen P.E.J., 2000, MNRAS 318, 889
\bibitem[2004]{Zanni04}
Zanni C., Murante G., Bodo G., Massaglia S., Rossi P., Ferrari A., 2005, A\&A 429, 399
\bibitem[\protect\citeauthoryear{Zaritsky, Gonzalez \& Zabludoff}{2004}]{Z.04}
Zaritsky, D., Gonzalez, A.H, \& Zabludoff, A.I., 2004, ApJ, 613, L93


\end{thebibliography}
\end{document}